\documentclass[11pt, draftclsnofoot, onecolumn]{IEEEtran}
\usepackage{pifont}
\usepackage{cite}
\usepackage[dvips]{epsfig}
\usepackage{graphicx}
\usepackage{calligra}
\usepackage[T1]{fontenc}
\usepackage{bbold}
\usepackage{mathrsfs}

\usepackage{subfigure}
\usepackage{color}
\usepackage{amsmath}
\usepackage{cases}

\usepackage{amssymb}
\usepackage[justification=centering]{caption}
\usepackage{framed}

\usepackage{subfigure}
\usepackage{color}
\usepackage{amsmath}
\usepackage{bm}
\usepackage{amssymb}
\usepackage{amsbsy}
\usepackage{bbm}
\usepackage{dsfont}
\usepackage[normalem]{ulem}
\usepackage{xcolor,cancel}
\usepackage{enumerate}
\usepackage{paralist}

\usepackage{algorithm, algorithmic}

\usepackage{amsthm}

\newtheorem{theorem}{Theorem}

\newtheorem{condition}{Condition}

\newtheorem{corollary}{Corollary}

\newtheorem{definition}{Definition}

\newtheorem{proposition}{Proposition}

\newtheorem{lemma}{Lemma}

\newtheorem{remark}{Remark}

\newtheorem{claim}{Claim}

\newtheorem{assumption}{Assumption}

\DeclareMathOperator{\esssup}{ess\,sup}

\newcommand{\bH}{{\bf{H}}}

\newcommand{\bbE}{\mathbbm{E}}

\newcommand{\bbP}{\mathbbm{P}}

\newcommand{\cF}{\mathcal{F}}

\newcommand{\cA}{\mathcal{A}}

\newcommand{\bone}{\mathbbm{1}}

\newcommand{\bx}{{\bf{x}}}

\newcommand{\btheta}{{\boldsymbol{\theta}}}

\newcommand{\bmu}{{\boldsymbol{\mu}}}

\newcommand{\tLambda}{{\widetilde{{\Lambda}}}}

\newcommand{\ignore}[1]{{}}

\newcounter{parentalgorithm}

\ifCLASSINFOpdf
\else
\fi

\begin{document}
\title{Quickest Detection of Time-Varying False Data Injection Attacks in Dynamic Linear Regression Models}
%\title{Quickest Detection of time-varying False Data Injection Attacks in Dynamic Smart Grid: Computational Efficient Algorithms and Analyses}

\author{{Jiangfan Zhang,~\IEEEmembership{Member,~IEEE},  Xiaodong Wang,~\IEEEmembership{Fellow,~IEEE}
}

\thanks{J. Zhang is with the Department of Electrical and Computer Engineering, Missouri University of Science and Technology, Roll, MO 65409 USA (e-mail: jiangfanzhang@mst.edu). X. Wang is with the Department of Electrical Engineering, Columbia University, New York, NY 10027 USA (e-mail:  wangx@ee.columbia.edu)}

%\thanks{This work was supported by the U. S. Army Research Laboratory and the U. S. Army Research Office and was accomplished under Agreement Numbers W911NF-14-1-0245 and W911NF-14-1-0261. The views and conclusions contained in this document are those of the authors and should not be interpreted as representing the official policies, either expressed or implied, of the Army Research Laboratory, Army Research Office, or the U.S. Government. The U.S. Government is authorized to reproduce and distribute reprints for Government purposes notwithstanding any copyright notation here on.}
%\thanks{J. Zhang and R. S. Blum are with the Department of Electrical and Computer Engineering, Lehigh University, Bethlehem, PA 18015 USA (e-mail: jiangfanzhang@lehigh.edu;  rblum@eecs.lehigh.edu).} 
%\thanks{Z. Chen and W. Xu are with the Department of Information Science and Electronic Engineering, Zhejiang University, Hangzhou, 310027 China (e-mail: joywithjoy@zju.edu.cn; wxu@zju.edu.cn).}

}


\maketitle

\begin{abstract}

Motivated by the sequential detection of false data injection attacks (FDIAs) in a dynamic smart grid, we consider a more general problem of sequentially detecting time-varying FDIAs in dynamic linear regression models. The unknown time-varying parameter vector in the linear regression model and the FDIAs impose a significant challenge for designing a computationally efficient detector. We first propose two Cumulative-Sum-type algorithms to address this challenge. One is called generalized Cumulative-Sum (GCUSUM) algorithm, and the other one is called relaxed generalized Cumulative-Sum (RGCUSUM) algorithm, which is a modified version of the GCUSUM. It can be shown that the computational complexity of the proposed RGCUSUM algorithm scales linearly with the number of observations. Next, considering Lordon's setup, for any given constraint on the expected false alarm period, a lower bound on the threshold employed in the proposed RGCUSUM algorithm is derived, which provides a guideline for the design of the proposed RGCUSUM algorithm to achieve the prescribed performance requirement. In addition, for any given threshold employed in the proposed RGCUSUM algorithm, an upper bound on the expected detection delay is also provided. The performance of the proposed RGCUSUM algorithm is also numerically studied in the context of an IEEE standard power system under FDIAs.

\end{abstract}




\section{Introduction}
\label{Section_Intro}

With the recent developments in sensing, signal processing, control, and communication, the smart grid system is tightly integrated with  cyber infrastructures, such as computer and communication networks, which makes it vulnerable to hostile cyber threats \cite{cui2012coordinated}.
The cyber-security and safety is a critical issue in smart grid systems since any outage, failure or cyber-attacks may bring about catastrophic consequences such as significant financial losses and blackouts in large geographic areas. When compared to detecting random errors, the detection of malicious data attacks in smart grid systems is far more difficult and is still an open problem, since  adversaries can judiciously choose the site of attack  and design attack data in a very sophisticated way. Hence, cyber-security in smart grid systems has drawn significant interest in recent years \cite{cui2012coordinated, huang2012state, kosut2011malicious, huang2011defending, huang2016real}.

In a smart grid system, meter readings  are periodically collected and stored by a supervisory control and data acquisition (SCADA) system. The meter measurements are utilized at a control center to estimate state variables of the smart grid system, such as bus voltages and phase angles, and then the operation of the smart grid system is performed and controlled based on these estimated states. If any adversary is able to falsify the meter measurements, the control center may produce erroneous state estimates, which brings about wrong decisions on billing, power dispatch, and even blackout. To this end, protection mechanism is of paramount importance  in  modern smart grid systems which should have the capability of detecting malicious attacks as quickly as possible. 
% modern smart grid systems must incorporate the protection mechanism, which has the capability of detecting malicious attacks as quickly as possible. 

Most of the existing works on attack detection in smart grid systems consider fixed-sample-size schemes, which aim to reach a satisfactory balance between the false alarm probability and detection probability. These fixed-sample-size schemes are useful when there is no strict latency constraint. However,  the detection of malicious attacks in smart grid systems is subject to strict latency constraints, since such attacks can be judiciously designed, and if not promptly detected, can cause increasingly catastrophic consequences to the system as time goes by. 
%In light of this, it is of paramount importance to minimize the detection delay after the presence of such attacks. 
The  sequential change detection methodology (also known as quickest detection), which minimizes
the expected detection delay subject to certain performance constraint on the average false alarm period, enables online  monitoring for smart
grid systems, and therefore suits well to attack detection in such systems.

\subsection{Summary of Results and Main Contributions}

We assume that every meter sequentially transmits its measurements to a control center. However the measurements from a subset of meters are corrupted by additive malicious data. Such attacks are referred to as false data injection attacks (FDIAs), which are considered as one of most detrimental attacks to smart grid systems \cite{cui2012coordinated}. The set of attacked meters and the malicious data injected into the attacked meters are unknown to the control center; and moreover, we assume that they can be distinct at different time instants. In addition, we assume that the state variables of the smart grid system are time-varying, and also unknown to the control center. It is worth mentioning that we do not assume knowledge of the state evolution dynamic, i.e.,  how the state variables change over time.  The control center aims at detecting any FDIAs as soon as possible when they are launched, and the sequential change detection scheme is considered in this paper.

The task of sequentially detecting time-varying FDIAs in such a dynamic system seems difficult. The reasons are mainly twofold. On one hand, since the state variables are time-varying and unknown to the control center, even though the control center observes a change in some statistic of the measurements, it is not clear  whether the change of the statistic is brought about by the unknown change of the state variables or by the occurrence of FDIAs. On the other hand, the number of possible sets of attacked meters is an exponential function of the number of meters. Thus, when the number of meters is large, which is typically the case in practice, it is strenuous to figure out which meters are attacked due to a large number of possibilities.

Motivated by the sequential FDIA detection task described above,
% this quickest attack detection problem in a dynamic smart grid system, 
in this paper, we consider a general problem of sequentially detecting  time-varying FDIAs in dynamic linear regression models, which subsumes the quickest FDIA detection in smart grid systems when the direct current (DC) power flow model is adopted.

Since the parameter vector in the dynamic linear regression model is unknown, we first resort to the generalized Cumulative Sum (CUSUM) algorithm. However, it can be shown that the computational complexity of the generalized CUSUM method scales exponentially with the number of observations, and therefore is infeasible in practice. 
To address this, we next propose a new CUSUM-type algorithm based on the generalized likelihood ratio (GLR), which is called relaxed generalized CUSUM (RGCUSUM) algorithm. The proposed RGCUSUM algorithm is computationally efficient and also robust to arbitrary time-varying parameter vector in the linear regression model and the FDIAs. To be specific, it can be shown that the computational complexity of the RGCUSUM algorithm just scales  linearly with the number of meters, and therefore, the RGCUSUM algorithm is more amenable than the generalized CUSUM to implementation in practice. Moreover,  this paper lays more emphasis on the performance characterization of the RGCUSUM algorithm with the aim of providing provable detection performance guarantee. In particular, considering Lordon's setup \cite{lorden1971procedures}, for any given constraint on the expected false alarm period, a lower bound on the threshold employed in the proposed algorithm is derived, which provides a guideline for the design of the RGCUSUM algorithm to achieve the prescribed performance requirement. In addition, for any given threshold employed in the proposed RGCUSUM algorithm, an upper bound on the expected detection delay is also provided.  In the numerical results, the performance of the proposed RGCUSUM algorithm is scrutinized in the context of an IEEE standard power system under FDIAs, and is compared with some representative algorithm. Numerical studies illustrate the reliable and quick response of the proposed RGCUSUM algorithm against time-varying FDIAs in dynamic linear regression models.

\subsection{Related Work}
\label{Subsection_Related_Work}

There is an increasing concern about the security of smart grid networks, see \cite{cui2012coordinated, huang2012state, kosut2011malicious, huang2011defending, huang2016real, liu2014local, yuan2011modeling, zhang2017functional, liang2017review, wang2013cyber, yan2012survey, zhang2018approaches} and
the references therein. Most of these works investigate the attack detection problem in smart grid systems under the fixed-sample-size
scheme, while this paper focuses on the sequential attack detection scheme. The quickest detection of time-varying FDIA in dynamic smart grid systems with the DC power flow model adopted has attracted more and more attention in recent years.

An adaptive CUSUM algorithm is proposed in \cite{huang2011defending}, which assumes that the state variables of the smart grid follows a Gaussian prior with some known mean and covariance, and also assumes that the FDIA is always positive and small so that the first-order approximation to the decision statistic is valid. The approach proposed in \cite{huang2011defending}, as a result, becomes inefficient for large FDIA, which could be more devastating to the system, and also for negative FDIA. On the contrary, we do not assume any prior for the state variables, and moreover, we do not impose any restriction on the sign and the magnitude of the FDIA. Real-time FDIA detection in static smart grid systems is studied in \cite{huang2016real}, where a CUSUM-type algorithm is proposed based on the residuals. In \cite{huang2016real}, the Rao test statistic is utilized to construct the decision statistic. However in many cases, such decision statistic cannot be evaluated due to the singularity of the covariance of the residuals, see Section \ref{Section_Application_Smart_Grid} for instance.  The quickest detection of FDIA in smart grids is also investigated in \cite{li2015quickest}, where the attack model is different from the model considered in this paper. To be specific, the set of effective attacked meters is assumed to be fixed over time in \cite{li2015quickest}, while we allow the adversaries to effectively attack distinct meters at different time instants. Thus, the attack model considered in this paper is more general. Moreover, no performance analysis of the proposed algorithm is provided in \cite{li2015quickest}, while in this paper, we not only propose a computationally efficient algorithm but also provide provable detection performance guarantees for the proposed algorithm. More recently, in \cite{kurt2018distributed}, a CUSUM-type algorithm based on the Kalman filter is proposed, which relies on the assumption that the state variables of the smart grid evolve over time by following a known linear model. On the contrary, in this paper,  no assumption is made on the dynamic of the time-varying state variables. In addition, no performance analysis is provided for the proposed algorithm in  \cite{kurt2018distributed}.

The remainder of the paper is organized as follows. The problem statement and background are described in Section \ref{Section_System_Model}. The generalized CUSUM algorithm and the computationally efficient RGCUSUM algorithm are introduced in Section \ref{Section_RGCUSUM}. In
Section \ref{Section_Performance_Analysis}, the performance of the proposed algorithms is investigated. In Section \ref{Section_Application_Smart_Grid}, numerical results are provided to illustrate the performance
of the proposed approach. Finally, Section \ref{Section_Conclusions} provides our conclusions.

\section{Problem Statement and Background}
\label{Section_System_Model}

In this section, we first formulate the general FDIA detection problem in dynamic linear regression models, which is motivated by the FDIA detection problem in dynamic smart grid systems where the DC power flow model is adopted. Then, the quickest detection technique for sequentially detecting the occurrence of attacks is briefly introduced.

% false data injection attacks in the dynamic direct current (DC) power flow model of smart grids to motivate the general data injection attack detection problems in dynamic regression models. Then, the quickest detection technique for sequentially detecting the occurrence of attacks is briefly introduced.

\subsection{Problem Statement}

We consider a dynamic linear regression model in which
\begin{equation} \label{regression_model}
{{\bf{x}}^{(t)}} = {\bf{H}}{{{\btheta }}^{(t)}} + {{\bf{n}}^{(t)}}
\end{equation}
where ${{\bf{x}}^{(t)}} \in {\mathbbm{R}}^M$, ${{{\btheta }}^{(t)}}\in {\mathbbm{R}}^N$ and ${{\bf{n}}^{(t)}}\in {\mathbbm{R}}^M$ are an $M$-dimensional vector of observations, an $N$-dimensional vector of unknown parameters  and an $M$-dimensional vector of noise at time instant $t$, respectively. ${\bH} \in {\mathbbm{R}}^{ M\times N}$ is an $M \times N$ linear model matrix. Typically, the number of measurements is greater than that of the unknown parameters in order to provide necessary redundancy against the noise effect, i.e., $M > N$.

Suppose that at time $t_a$, a malicious attacker intentionally manipulates the observation vector by injecting a sequence of unknown false data $\{{\bf b}^{(1)}, {\bf b}^{(2)}, {\bf b}^{(3)},...\}$. Accordingly, we write the pre- and post-attack observations as
\begin{equation} \label{problem_formulation}
\left\{ \begin{array}{l}
{{\bf{x}}^{(t)}} = {\bf{H}}{\btheta ^{(t)}} + {{\bf{n}}^{(t)}} \qquad \quad \text{if } t< t_a,\\
{{\bf{x}}^{(t)}} = {\bf{H}}{\btheta ^{(t)}} + {{\bf a}^{(t)}} + {{\bf{n}}^{(t)}} \quad \text{if } t \ge t_a,
\end{array} \right.
\end{equation}
where ${\bf a}^{(t_a + t-1)} = {\bf b}^{(t)}$ for any $t_a$ and $t \ge 1$.
Note that the injected false data ${\bf a}^{(t)}$ can be decomposed into two parts 
\begin{equation}
{{\bf{a}}^{(t)}} = {\bf{H}}{{\bf{c}}^{(t)}} + {\bmu ^{(t)}}
\end{equation}
where ${\bf{H}}{{\bf{c}}^{(t)}}$ denotes the component of ${\bf a}^{(t)}$ that lies in the column space of $\bf H$, while ${\bmu ^{(t)}}$ represents the component of ${\bf a}^{(t)}$ that lies in the complementary space ${{\cal R}^ \bot({\bf H}) }$ of the column space of $\bf H$, that is
\begin{equation} \label{Define_mu}
{\bmu ^{(t)}}  = {\bf{P}}_{\bf{H}}^ \bot {{\bf{a}}^{(t)}} \in {{\cal R}^ \bot({\bf H}) }
\end{equation}
where
\begin{equation} \label{Define_P_H_per}
{\bf{P}}_{\bf{H}}^ \bot  \triangleq {\bf{I}} - {\bf{H}}{\left( {{{\bf{H}}^T}{\bf{H}}} \right)^{ - 1}}{{\bf{H}}^T}.
\end{equation}
As illustrated in \cite{liu2011false}, ${\bmu ^{(t)}}$ is the only informative part of the injected false data that is detectable. The reason is that since the parameter vector ${\btheta ^{(t)}}$ is unknown, the other part ${\bf{H}}{{\bf{c}}^{(t)}}$ of ${\bf a}^{(t)}$ is not distinguishable from ${\bf{H}}{\btheta ^{(t)}}$, and hence can bypass any monitoring system.

Let $\rho_L$ and $\rho_U$ denote the lower and upper bounds on the magnitudes of the nonzero elements of ${\bmu ^{(t)}}$, respectively. Let ${\cal A}^{(t)}$ represent the set of nonzero elements of ${\bmu ^{(t)}}$ at time instant $t$. As such, we can rewrite the pre- and post-attack model as 
\begin{equation} \label{mu_constraint}
\begin{array}{*{20}{l}}
{t < {t_a}:}&{\mu _m^{(t)} = 0,m = 1,2,...,M,}\\
{t \ge {t_a}:}&{\left\{ \begin{array}{l}
	{\rho _L} \le |\mu _m^{(t)}| \le {\rho _U},m \in {{\cal A}^{(t)}},\\
	\mu _m^{(t)} = 0,m \notin {{\cal A}^{(t)}},
	\end{array} \right.}
\end{array}
\end{equation}
%\begin{equation} \label{mu_constraint}
%\left\{ \begin{array}{l}
%\mu _m^{(t)} = 0, \quad m = 1,2,...,M, \; \; t < {t_a},\\
%{\rho_L} \le |\mu _m^{(t)}| \le \rho_U , \quad m \in {\cal A}^{(t)}, \; \; t \ge {t_a},
%\end{array} \right.
%\end{equation}
where $\mu _m^{(t)}$ is the $m$-th element of ${\bmu ^{(t)}}$. It is worth mentioning that since ${\bmu ^{(t)}}$ can be time-varying, so can be the set ${\cal A}^{(t)}$, which is referred to as the set of attacked observations at time instant $t$ thereafter. Our goal is to detect the attacks as soon as possible after their occurrence at time $t_a$. 
The quickest detection technique, that exploits the statistical difference before and after $t_a$, provides a suitable framework to achieve this goal, and hence,  in this paper, we resort to these techniques to detect the false data injection attacks described in (\ref{problem_formulation}).

The FDIA detection problem described in (\ref{problem_formulation}) is motivated by the problem of detecting FDIAs in smart grid systems. To be specific, consider $M$ meters in an $(N + 1)$-bus smart grid system. Then the dynamic DC power flow model of the system can be exactly formulated as in (\ref{regression_model}) where the unknown parameter vector $\btheta^{(t)}$ is the $N$ phase angles (one reference angle) at time $t$ which evolves over time due to the time-varying workload in the system, and the vector of observations $\bx^{(t)}$ is the measurements of the power flows and power injections at the $M$ meters at time instant $t$. The linear model matrix ${\bf H}$ in the context of a smart grid is the measurement matrix which depends on the topology of the smart grid, the placement of the meters, and the susceptance of each transmission line. Moreover, the attack model in (\ref{problem_formulation}) essentially demonstrates that the adversaries employ a sequence of false data to corrupt the meter readings starting at some time instant. For more details about the application of the FDIA detection problem described in (\ref{problem_formulation}) to smart grid systems, please refer to Section \ref{Section_Application_Smart_Grid}.

If adversaries could perfectly know the measurement matrix ${\bf H}$ and manipulate the observations ${\bf x}^{(t)}$ at any meter they want, then they would be able to design the false data injection attack ${\bf a}^{(t)}$ to be stealth attack which lies in the column space  $\bf H$, i.e., $\bmu^{(t)}={\bf 0}$, and hence bypass any security system \cite{liu2011false}. Fortunately, this is not the case in a smart grid system. In particular, a smart grid system is large scale in practice, and the meters in the system are widely distributed. Hence, the adversaries typically can only access and manipulate a subset of meters which are close to them due to their limited amount of resources. Moreover, by securing sufficient number of meters in the smart grid system,  stealth attacks can be prevented according to \cite{kosut2011malicious}, where the fundamental limit of the stealth attacks were studied. Therefore, we assume that if the system is under attack, then ${\bmu ^{(t)}} \ne {\bf 0}$. The constant ${\rho_L}$ indicates the lower bound on the nonzero element of ${\bmu ^{(t)}}$ that draws security concerns, and the constant $\rho_U$ represents the limited power of the adversaries. In addition, we assume that $\{{{\bf{n}}^{(t)}}\}$ in (\ref{problem_formulation}) is a sequence of independent and identically distributed (i.i.d.) noise vectors obeying Gaussian distribution ${\cal N}({\bf{0}},{\sigma ^2}{{\bf{I}}_M})$ with mean $\bf 0$  and covariance $\sigma^2{\bf{I }}_M$.

\subsection{Quickest Detection}

In quickest detection, observations are made sequentially in time, and then stop when a change is declared. The sequential change detector aims at minimizing the expected detection delay after the change-point. The commonly used performance measure, proposed by Lorden, is the worst-case expected detection delay which is defined as \cite{lorden1971procedures}
\begin{equation} \label{lorden_formulation}
J\left( T \right) \buildrel \Delta \over = \mathop {\sup }\limits_{t_a}  {J_{{t_a}}}\left( {{T}} \right),
\end{equation}
with
\begin{equation} \label{Define_J_t_a}
%{J_{{t}}}\left( {{T}} \right) \buildrel \Delta \over = {\esssup _{{{\cal F}_{{t} - 1}}}} {\bbE_{{t}}}\left\{ {\left. {{{\left( {{T} - {t} + 1} \right)}^ + }} \right|{{\cal F}_{{t} - 1}}} \right\},
{J_{{t_a}}}\left( {{T}} \right) \buildrel \Delta \over = {\esssup _{{{\cal F}_{{t_a} - 1}}}} {\bbE_{{t_a}}}\left\{ {\left. {{{\left( {{T} - {t_a} + 1} \right)}^ + }} \right|{{\cal F}_{{t_a} - 1}}} \right\}
\end{equation}
where the random variable $T$ is a stopping time corresponding to a certain sequential detection scheme and $\cF_{t_a}$ is the filtration generated by all the observations up to time $t_a$. 
The expectation ${\bbE_{t_a}}$ is evaluated with respect to the true distribution of ${\bf x}^{(1)}$, ${\bf x}^{(2)}$,  ${\bf x}^{(3)}$, ... when the change occurs at time instant $t_a$.
%the post-change probability measure conditioned on the change-point $t$ and the all observations up to time $t$. 
The essential supremum in (\ref{Define_J_t_a}) is obtained over ${{\cal F}_{t_a-1}}$, yielding the least favorable situation for the expected detection delay. The supremum in (\ref{lorden_formulation}) is obtained over $t_a$, implying that the change occurs at such a point that the expected detection delay is maximized. To summarize, $J(T)$ characterizes the expected detection delay for the worst possible change point and the worst possible history of observations before the change point. While  small expected detection delay under attack brings about timely alarmed reaction, the running length under no attack needs to be guaranteed to be large enough to avoid frequent false alarms. To this end, the sequential change detection problem is formulated as follows:
\begin{equation} \label{changepoint_detection_problem_formulation}
\mathop {\inf }  \limits_T   J\left( T \right)  {\text{  subject to  }}   {\bbE_\infty }\left\{ T \right\} \ge \gamma.
\end{equation}
Note that  the expectation ${\bbE_{\infty}}$ is evaluated with respect to the probability measure where $t_a=\infty$, i.e., no change occurs, and $\gamma$ is a prescribed constant which specifies the required lower bound on the expected false alarm period. To proceed with our FDIA detection problem, we denote the pre-attack and post-attack probability density functions of the observation ${\bf x}^{(t)}$ as ${f_{u}}( {{{\bf{x}}^{(t)}}\left| {{\btheta ^{(t)}}} \right.} )$ and ${f_{a}}( {{{\bf{x}}^{(t)}}\left| {{\btheta ^{(t)}},{{\bf{a}}^{(t)}}} \right.} )$, respectively. If all the parameters $\{\btheta^{(t)}\}$ and $\{{\bf a}^{(t)}\}$ were known, the quickest detection problem in (\ref{changepoint_detection_problem_formulation}) is optimally solved by the well-known CUSUM test \cite{moustakides1986optimal}
\begin{equation} \label{cusum_test}
{T_{{\rm{CUSUM}}}} = \min \left\{ {K:\mathop {\max }\limits_{1 \le k \le K} \sum\limits_{t = k}^K {\ln \frac{{{f_a}\left( {{{\bf{x}}^{(t)}}\left| {{\btheta ^{(t)}},{{\bf{a}}^{(t)}}} \right.} \right)}}{{{f_u}\left( {{{\bf{x}}^{(t)}}\left| {{\btheta ^{(t)}}} \right.} \right)}} \ge h} } \right\},
\end{equation}
where the threshold $h$ is determined by the constraint in (\ref{changepoint_detection_problem_formulation}). For a given $K$, the value of $k$ which maximizes the test statistic in (\ref{cusum_test}) can be considered as an estimate of change point. It is well-known that the nonlinear accumulation of the log-likelihood ratios in (\ref{cusum_test}) can be written in a recursive way, and hence can be easily implemented in practice with low complexity \cite{basseville1993detection, tartakovsky2014sequential}. 

Note that in our FDIA detection problem, the parameters $\{\btheta^{(t)}\}$ and $\{{\bf a}^{(t)}\}$ are unknown, rendering the CUSUM test infeasible.
% and need to be estimated, as estimating $\theta$ is the essential task of the network. 
To address this, this paper resort to the generalized likelihood ratio (GLR) method by replacing the unknown parameters with their maximum likelihood estimates (MLE) \cite{basseville1993detection, tartakovsky2014sequential}. We next propose a sequential attack detector based on the GLR idea.

\section{Sequential Attack Detection Based on Relaxed Generalized CUSUM Test}
\label{Section_RGCUSUM}

In this section, we first derive a GLR-based FDIA sequential detector, namely generalized CUSUM (GCUSUM), but its complexity grows exponentially with respect to the size $M$ of the observation vector, rendering the GCUSUM infeasible when $M$ is large. Then, we propose a modified sequential detector whose computational complexity grows linearly with respect to $M$.

\subsection{Generalized CUSUM Test}
\label{Section_GCUSUM}
%
%Let $\kappa $ denote the minimum distortion which draws security attentions and let ${\cal A}^{(t)}$ denote the set of meters where the injected false data in the space $\cR^{\perp}(\bf H)$ is larger than the minimum distortion at time instant $t$, that is, 
%\begin{equation} \label{mu_constraint}
%\left\{ \begin{array}{l}
%\mu _m^{(t)} = 0, \quad m = 1,2,...,M, \; \; t < {t_a},\\
%{\rho_L} \le |\mu _m^{(t)}| \le \rho_U , \quad m \in {\cal A}^{(t)}, \; \; t \ge {t_a}.
%\end{array} \right.
%\end{equation}

Based on the model in (\ref{mu_constraint}), by replacing the unknown parameters $\{\btheta^{(t)}\}$ and $\{{\bf a}^{(t)}\}$ in (\ref{cusum_test}) with their MLEs, the GCUSUM test can be written as
\begin{equation} \label{GCUSUM_test}
{T_G} = \min \left\{ {K:\mathop {\max }\limits_{1 \le k \le K} \mathop {\sup }\limits_{\{{\cal A}^{(t)}\}} \Lambda _{k}^{(K)} \ge h} \right\},
\end{equation}
where 
%the statistic $\Lambda _{k}^{(K)}$ is defined as
\begin{align} \notag
\Lambda _{k}^{(K)} & \buildrel \Delta \over =  \ln \frac{{\mathop {\sup }\limits_{ {{\btheta ^{(t)}},{{\bf{a}}^{(t)}}:{{\{ {\rho_L} \le |\mu _m^{(t)}| \le \rho_U \} }_{m \in {\cal A}^{(t)}}},{\bmu ^{(t)}} \in {{\cal R}^ \bot }\left( {\bf{H}} \right)} } \prod\limits_{t = 1}^{k - 1} {{f_u}\left( {{{\bf{x}}^{(t)}}\left| {{\btheta ^{(t)}}} \right.} \right)} \prod\limits_{t = k}^K {{f_a}\left( {{{\bf{x}}^{(t)}}\left| {{\btheta ^{(t)}},{{\bf a} ^{(t)}}} \right.} \right)} }}{{\mathop {\sup }\limits_{{\btheta ^{(t)}}} \prod\limits_{t = 1}^K {{f_u}\left( {{{\bf{x}}^{(t)}}\left| {{\btheta ^{(t)}}} \right.} \right)} }}\\  \notag
& = \ln \frac{{\prod\limits_{t = 1}^{k - 1} {\mathop {\sup }\limits_{{\btheta ^{(t)}}} {f_u}\left( {{{\bf{x}}^{(t)}}\left| {{\btheta ^{(t)}}} \right.} \right)} \prod\limits_{t= k}^K {\mathop {\sup }\limits_{{{\btheta ^{(t)}},{{\bf{a}}^{(t)}}:{{\{ {\rho_L} \le |\mu _m^{(t)}| \le \rho_U \} }_{m \in {\cal A}^{(t)}}},{\bmu ^{(t)}} \in {{\cal R}^ \bot }\left( {\bf{H}} \right)}} {f_a}\left( {{{\bf{x}}^{(t)}}\left| {{\btheta ^{(t)}},{{\bf a} ^{(t)}}} \right.} \right)} }}{{\prod\limits_{t = 1}^K {\mathop {\sup }\limits_{{\btheta ^{(t)}}} {f_u}\left( {{{\bf{x}}^{(t)}}\left| {{\btheta ^{(t)}}} \right.} \right)} }}\\ \label{Lambda_k_K}
%& = \sum\limits_{t = k}^K \Lambda _{k,t}^{(K)},
& = \sum\limits_{t = k}^K \! {\Bigg[ \! {\underbrace {\mathop {\sup }\limits_{ {{\btheta ^{(t)}},{{\bf{a}}^{(t)}}:{{\{ {\rho_L} \le |\mu _m^{(t)}| \le \rho_U \} }_{m \in {\cal A}^{(t)}}},{\bmu ^{(t)}} \in {{\cal R}^ \bot }\left( {\bf{H}} \right)} } \ln {f_a}\left( {{{\bf{x}}^{(t)}}\left| {{\btheta ^{(t)}},{{\bf a} ^{(t)}}} \right.} \right) \! -\! \mathop {\sup }\limits_{{\btheta ^{(t)}}} \ln {f_u}\left( {{{\bf{x}}^{(t)}}\left| {{\btheta ^{(t)}}} \right.} \right)}_{ \buildrel \Delta \over = \Lambda _{k,t}^{(K)}}} \Bigg]}.
\end{align}
%and ${{f_u}\left( {{{\bf{x}}^{(t)}}\left| {{\btheta ^{(t)}}} \right.} \right)}$ and ${{f_a}\left( {{{\bf{x}}^{(t)}}\left| {{\btheta ^{(t)}},{{\bf a} ^{(t)}}} \right.} \right)}$ are the pre-attack and post-attack probability density functions of ${{\bf{x}}^{(t)}}$, respectively.
Given that ${\bf n}^{(t)} \mathop  \sim \limits^{i.i.d.} {\cal N}({\bf{0}},{\sigma ^2}{{\bf{I}}_M})$, $\Lambda _{k,t}^{(K)}$ can be simplified to 
%$\{{\bf n}^{(t)}\}$ is a sequence of i.i.d. white Gaussian noise with zero mean and covariance $\sigma^2{\bf{I }}_M$, $\Lambda _{k,t}^{(K)}$ can be simplified to 
\begin{align} \notag
\Lambda _{k,t}^{(K)} 
%& \triangleq  \mathop {\sup }\limits_{{{\btheta ^{(t)}},{{\bf{a}}^{(t)}}:{{\{ {\rho_L} \le |\mu _m^{(t)}| \le \rho_U \} }_{m \in {\cal A}^{(t)}}},{\bmu ^{(t)}} \in {{\cal R}^ \bot }\left( {\bf{H}} \right)}} \ln {f_a}\left( {{{\bf{x}}^{(t)}}\left| {{\btheta ^{(t)}},{{\bf a} ^{(t)}}} \right.} \right) - \mathop {\sup }\limits_{{\theta ^{(t)}}} \ln {f_u}\left( {{{\bf{x}}^{(t)}}\left| {{\btheta ^{(t)}}} \right.} \right)\\ \notag
&  = \mathop {\sup }\limits_{ {{\btheta ^{(t)}},{{\bf{a}}^{(t)}}:{{\{ {\rho_L} \le |\mu _m^{(t)}| \le \rho_U \} }_{m \in {\cal A}^{(t)}}},{\bmu ^{(t)}} \in {{\cal R}^ \bot }\left( {\bf{H}} \right)} }  - \frac{1}{{2{\sigma ^2}}}{\left( {{{\bf{x}}^{(t)}} - {\bf{H}}{\btheta ^{\left( t \right)}} - {{\bf a} ^{(t)}}} \right)^T}\left( {{{\bf{x}}^{(t)}} - {\bf{H}}{\btheta ^{\left( t \right)}} - {{\bf a} ^{(t)}}} \right)  \\ \notag
& \qquad - \mathop {\sup }\limits_{{\btheta ^{(t)}}}  - \frac{1}{{2{\sigma ^2}}}{\left( {{{\bf{x}}^{(t)}} - {\bf{H}}{\btheta ^{(t)}}} \right)^T}\left( {{{\bf{x}}^{(t)}} - {\bf{H}}{\btheta ^{(t)}}} \right) \\ \notag
& = \mathop {\sup }\limits_{ {{{\bf{a}}^{(t)}}:{{\{ {\rho_L} \le |\mu _m^{(t)}| \le \rho_U \} }_{m \in {\cal A}^{(t)}}},{\bmu ^{(t)}} \in {{\cal R}^ \bot }\left( {\bf{H}} \right)} }  - \frac{1}{{2{\sigma ^2}}}\left[ {{{\left( {{{\bf{x}}^{(t)}} - {{\bf a} ^{(t)}}} \right)}^T}{\bf{P}}_{\bf{H}}^ \bot \left( {{{\bf{x}}^{(t)}} - {{\bf a} ^{(t)}}} \right) - {{\left( {{{\bf{x}}^{(t)}}} \right)}^T}{\bf{P}}_{\bf{H}}^ \bot {{\bf{x}}^{(t)}}} \right]\\  \notag
%& = \mathop {\sup }\limits_{ {{{\bf{a}}^{(t)}}:{{\{ {\rho_L} \le |\mu _m^{(t)}| \le \rho_U \} }_{m \in {\cal A}^{(t)}}},{\bmu ^{(t)}} \in {{\cal R}^ \bot }\left( {\bf{H}} \right)} } \frac{1}{{2{\sigma ^2}}}\left[ {2{{\left( {{{\bf a} ^{(t)}}} \right)}^T}{\bf{P}}_{\bf{H}}^ \bot {{\bf{x}}^{(t)}} - {{\left( {{{\bf a} ^{(t)}}} \right)}^T}{\bf{P}}_{\bf{H}}^ \bot {{\bf a} ^{(t)}}} \right]\\ \notag
&  = \mathop {\sup }\limits_{ {{{\bmu}^{(t)}}:{{\{ {\rho_L} \le |\mu _m^{(t)}| \le \rho_U \} }_{m \in {\cal A}^{(t)}}},{\bmu ^{(t)}} \in {{\cal R}^ \bot }\left( {\bf{H}} \right)} } \frac{1}{{2{\sigma ^2}}}\left[ {2{{\left( {{\bmu ^{(t)}}} \right)}^T}{{{\bf{\tilde x}}}^{(t)}} - \left\| {{\bmu ^{(t)}}} \right\|_2^2} \right] \\ \label{LLR_single_simplified}
& = \mathop {\sup }\limits_{ {{{\bmu}^{(t)}}:{{\{ {\rho_L} \le |\mu _m^{(t)}| \le \rho_U \} }_{m \in {\cal A}^{(t)}}},{\bmu ^{(t)}} \in {{\cal R}^ \bot }\left( {\bf{H}} \right)} } \frac{1}{{2{\sigma ^2}}}\sum\limits_{m \in {{\cal A}^{(t)}}} {\left[ {2\mu _m^{(t)}{\tilde x_m^{(t)}} - {{\left( {\mu _m^{(t)}} \right)}^2}} \right]} 
%& = \frac{1}{{2{\sigma ^2}}}\sum\limits_{m \in {{\cal A}^{(t)}}} {\mathop {\sup }\limits_{ {{{\bmu}^{(t)}}:{{\{ {\rho_L} \le |\mu _m^{(t)}| \le \rho_U \} }_{m \in {\cal A}^{(t)}}},{\bmu ^{(t)}} \in {{\cal R}^ \bot }\left( {\bf{H}} \right)} } \left[ {2\mu _m^{(t)}{\tilde x_m^{(t)}} - {{\left( {\mu _m^{(t)}} \right)}^2}} \right]}
\end{align}
by employing (\ref{Define_mu}). In (\ref{LLR_single_simplified}),  ${\bf{P}}_{\bf{H}}^ \bot$ is defined in (\ref{Define_P_H_per}),   ${{{\bf{\tilde x}}}^{(t)}}$ is the component of ${\bf x}^{(t)}$ in the complementary space of the column space of $\bf H$, i.e.,
%\cbr(Only using ${{{\bf{\tilde x}}}^{(t)}}$ instead of ${\bf x}^{(t)}$ to allow the upper bound of the test statistic) \cbk
%${\bf{P}}_{\bf{H}}^ \bot  \triangleq {\bf{I}} - {\bf{H}}{\left( {{{\bf{H}}^T}{\bf{H}}} \right)^{ - 1}}{{\bf{H}}^T}$, and (\ref{LLR_single_simplified}) is due to the fact that ${\bf{P}}_{\bf{H}}^ \bot {{\bf a} ^{(t)}} = {{\bf a} ^{(t)}}$.
\begin{equation} \label{tilde_x_define}
{{{\bf{\tilde x}}}^{(t)}} \buildrel \Delta \over = {\bf{P}}_{\bf{H}}^ \bot {{\bf{x}}^{(t)}},
\end{equation}
and ${\tilde x_m^{(t)}}$ and ${\mu_m^{(t)}}$ are the $m$-th elements of ${{{\bf{\tilde x}}}^{(t)}}$ and $\bmu^{(t)}$, respectively.

As a result, by employing (\ref{Lambda_k_K})  and (\ref{LLR_single_simplified}), at time instant $K$, the test statistic of the GCUSUM in (\ref{GCUSUM_test}) can be rewritten as
\begin{align} \notag
{V_K} & \buildrel \Delta \over = \mathop {\max }\limits_{1 \le k \le K} \mathop {\sup }\limits_{\{ {{\cal A}^{(t)}}\} } \Lambda _k^{(K)}\\ \notag
& = \mathop {\max }\limits_{1 \le k \le K} \mathop {\sup }\limits_{\{ {{\cal A}^{(t)}}\} } \sum\limits_{t = k}^K {\Lambda _{k,t}^{(K)}} \\ \label{Define_w_t}
& = \mathop {\max }\limits_{1 \le k \le K} \sum\limits_{t = k}^K {\underbrace {\mathop {\sup }\limits_{{{\cal A}^{(t)}}} \mathop {\sup }\limits_{{\bmu ^{(t)}}:{{\{ {\rho _L} \le |\mu _m^{(t)}| \le {\rho _U}\} }_{m \in {{\cal A}^{(t)}}}},{\bmu ^{(t)}} \in {{\cal R}^ \bot }\left( {\bf{H}} \right)} \frac{1}{{2{\sigma ^2}}}\sum\limits_{m \in {{\cal A}^{(t)}}} {\left[ {2\mu _m^{(t)}\tilde x_m^{(t)} - {{\left( {\mu _m^{(t)}} \right)}^2}} \right]} }_{ \buildrel \Delta \over = {v_t}}},
\end{align}
and further, can be calculated in a recursive way that
\begin{equation} \label{recursive_GCUSUM}
{V_K} = \mathop {\max }\limits_{1 \le k \le K} \sum\limits_{t = k}^K {{v_t}}  = \max \left\{ {\mathop {\max }\limits_{1 \le k \le K - 1} \sum\limits_{t = k}^K {{v_t}} ,{v_K}} \right\} = \max \left\{ {{V_{K - 1}},0} \right\} + {v_K}, \; \text{ with } V_0 =0.
\end{equation}
 We summarize the procedure for implementing the GCUSUM in Algorithm \ref{Algorithm_GCUSUM}.  The recursive expression in (\ref{recursive_GCUSUM}) implies that the test statistic of the GCUSUM can be efficiently obtained if $v_t$ is available for for $t$. 
However, the complexity of computing the statistic $v_t$ in (\ref{Define_w_t}) is generally very high, especially when $M$ is large. To be specific, it is seen from (\ref{recursive_GCUSUM}) that in order to obtain $v_t$, ${\frac{1}{{2{\sigma ^2}}}\sum_{m \in {{\cal A}^{(t)}}} {[ {2\mu _m^{(t)}\tilde x_m^{(t)} - {{( {\mu _m^{(t)}} )^2}}} ]} }$ needs to be maximized over $\bmu^{(t)}$ for a given ${\cal A}^{(t)}$ first, and then maximized over all possible ${\cal A}^{(t)}$. Since the number of possible ${\cal A}^{(t)}$ is on the order of $2^M$, $v_t$  may be computed exactly by exhaustively searching through all possible ${\cal A}^{(t)}$ for a small $M$, while for a large $M$, this is not feasible in practice. To this end, the GCUSUM test in (\ref{GCUSUM_test}) is not amenable to implementation in practice, which motivates us to pursue more computationally efficient algorithms.
\begin{algorithm}[htb]
	\caption{Generalized CUSUM Algorithm}
	\begin{algorithmic}[1]
		\STATE \textbf{Input}:  ${\bf x}^{(t)}$, $\bf H$, $\rho_L$, $\rho_U$, $\sigma^2$ and $h$
		\STATE  \textbf{Output}: $T_{{G}}$
		\STATE \textbf{Initialization}: $t \leftarrow 0$, $V_0 \leftarrow 0$
		\WHILE{$V_t <h$}
		\STATE $t \leftarrow t+1$
		\STATE Evaluate $\Lambda _{k,t}^{(K)}$ by using (\ref{LLR_single_simplified}). 
		\STATE $v_t \leftarrow \mathop {\sup }_{{{\cal A}^{(t)}}} \Lambda _{k,t}^{(K)}$
		\STATE Update the decision statistic: ${V_t} \leftarrow \max \left\{ {{V_{t - 1}},0} \right\} + {v_t}$.
		\ENDWHILE
		\STATE $T_{{G}}\leftarrow t$, declare an FDIA.
	\end{algorithmic}
	\label{Algorithm_GCUSUM}
\end{algorithm}

%\cbr Add the recursive expression of $\mathop {\max }\limits_{1 \le k \le K} \mathop {\sup }\limits_{\{{\cal A}^{(t)}\}} \Lambda _{k}^{(K)}$. \cbk
%
%\cbr No closed-form expression, it is hard to analyze the performance, difficult to compute in practice... \cbk

\subsection{Relaxed Generalized CUSUM Test}

In order to facilitate the computation of the test statistic of the GCUSUM in (\ref{recursive_GCUSUM}), we first simplify the computation of the statistic $\Lambda _{k,t}^{(K)}$ in (\ref{LLR_single_simplified}).

It is seen from (\ref{LLR_single_simplified}) that $\Lambda _{k,t}^{(K)}$ is the solution of a constrained optimization problem, and there is generally no closed-form expression for $\Lambda _{k,t}^{(K)}$  due to the constraints on $\bmu^{(t)}$ in (\ref{LLR_single_simplified}). Here, we relax the constraint ${\bmu ^{(t)}} \in {{\cal R}^ \bot }(\bf{H})$, and correspondingly, $\Lambda _{k,t}^{(K)}$ can be bounded from above as per
\begin{align} \notag
%& \mathop {\sup }\limits_{{\btheta ^{\left( t \right)}},{{\{ |\mu _m^{(t)}| \ge \kappa \} }_{m \in {{\cal A}^{(t)}}}}} \ln {f_a}\left( {{{\bf{x}}^{(t)}}\left| {{\btheta ^{\left( t \right)}},{{\bf a} ^{(t)}}} \right.} \right) - \mathop {\sup }\limits_{{\btheta ^{(t)}}} \ln {f_u}\left( {{{\bf{x}}^{(t)}}\left| {{\btheta ^{(t)}}} \right.} \right)\\ \notag
\Lambda _{k,t}^{(K)} &  = \mathop {\sup }\limits_{ {{{\bmu}^{(t)}}:{{\{ {\rho_L} \le |\mu _m^{(t)}| \le \rho_U \} }_{m \in {\cal A}^{(t)}}},{\bmu ^{(t)}} \in {{\cal R}^ \bot }\left( {\bf{H}} \right)} } \frac{1}{{2{\sigma ^2}}}\sum\limits_{m \in {{\cal A}^{(t)}}} {\left[ {2\mu _m^{(t)}{\tilde x_m^{(t)}} - {{\left( {\mu _m^{(t)}} \right)}^2}} \right]} \\ \notag
& \le \mathop {\sup }\limits_{ {{{\bmu}^{(t)}}:{{\{ {\rho_L} \le |\mu _m^{(t)}| \le \rho_U \} }_{m \in {\cal A}^{(t)}}} } } \frac{1}{{2{\sigma ^2}}}\sum\limits_{m \in {{\cal A}^{(t)}}} {\left[ {2\mu _m^{(t)}{\tilde x_m^{(t)}} - {{\left( {\mu _m^{(t)}} \right)}^2}} \right]} \\ \notag
%& = \frac{1}{{2{\sigma ^2}}}\sum\limits_{m \in {{\cal A}^{(t)}}} {\mathop {\sup }\limits_{ {{{\bmu}^{(t)}}:{{\{ {\rho_L} \le |\mu _m^{(t)}| \le \rho_U \} }_{m \in {\cal A}^{(t)}}},{\bmu ^{(t)}} \in {{\cal R}^ \bot }\left( {\bf{H}} \right)} } \left[ {2\mu _m^{(t)}{\tilde x_m^{(t)}} - {{\left( {\mu _m^{(t)}} \right)}^2}} \right]} \\ \notag
& = \frac{1}{{2{\sigma ^2}}}\sum\limits_{m \in {{\cal A}^{(t)}}} {\mathop {\sup }\limits_{ {{{\mu}_m^{(t)}}:{{\{ {\rho_L} \le |\mu _m^{(t)}| \le \rho_U \} }_{m \in {\cal A}^{(t)}}}  } } \left[ {2\mu _m^{(t)}{\tilde x_m^{(t)}} - {{\left( {\mu _m^{(t)}} \right)}^2}} \right]} \\ \notag
& = \sum\limits_{m \in {{\cal A}^{(t)}}} {\zeta _m^{(t)}} \\
& \triangleq \tLambda _{k,t}^{(K)}, 
\end{align}
where 
\begin{equation} \label{tilde_zeta_m_t_define}
\zeta _m^{(t)} \triangleq \left\{ \begin{array}{l}
\frac{1}{{2{\sigma ^2}}}{\left( {{\tilde x_m^{(t)}}} \right)^2} \; \qquad  \quad \qquad \; \; \text{if} \;\; {\rho_L} \le \left| {{\tilde x_m^{(t)}}} \right| \le  {\rho_U},\\
\frac{1}{{2{\sigma ^2}}}\left( {2\left| {{\tilde x_m^{(t)}}} \right|\rho_L  - {\rho_L ^2}} \right) \quad \text{if} \;\; \left| {{\tilde x_m^{(t)}}} \right| < {\rho_L},\\
\frac{1}{{2{\sigma ^2}}}\left( {2\left| {{\tilde x_m^{(t)}}} \right|\rho_U  - {\rho_U ^2}} \right) \quad \text{if} \;\; \left| {{\tilde x_m^{(t)}}} \right| > {\rho_U}.
\end{array} \right.
\end{equation}

%${\zeta _m^{(t)},0}$ depends on ${\tilde{x}}_m^{(t)}$, so we first exam the property of ${\tilde{x}}_m^{(t)}$.

%Let 
%\begin{equation} \label{P_pm}
%{\bf{P}}_{\bf{H}}^ \bot  = \left( \begin{array}{l}
%{\bf{p}}_1^T\\
%{\bf{p}}_2^T\\
%\vdots \\
%{\bf{p}}_M^T
%\end{array} \right)
%\end{equation}
%By employing the definition of ${\tilde {\bf x}^{(t)}}$ in (\ref{tilde_x_define}), we can obtain
%\begin{equation}
%\tilde x_m^{(t)} = {\bf{p}}_m^T{{\bf{x}}^{(t)}},
%\end{equation}
%which yields that the pre-attack and post-attack distributions of $\tilde x_m^{(t)}$ can be expressed as
%%\begin{equation} \label{tilde_x_distribution}
%%\tilde x_m^{(t)} \sim {\cal N}\left( {0,{\sigma ^2}\left\| {{{\bf{p}}_m}} \right\|_2^2} \right) \quad \text{ and } \quad \tilde x_m^{(t)} \sim {\cal N}\left( {{\bf{p}}_m^T{{\bf{a}}^{(t)}},{\sigma ^2}\left\| {{{\bf{p}}_m}} \right\|_2^2} \right),
%%\end{equation}
%\begin{equation} \label{tilde_x_distribution_0}
%\tilde x_m^{(t)} \sim {\cal N}\left( {0,{\sigma ^2}\left\| {{{\bf{p}}_m}} \right\|_2^2} \right)
%\end{equation}
%\begin{equation} \label{tilde_x_distribution_1}
%\text{and} \quad \tilde x_m^{(t)} \sim {\cal N}\left( {{\bf{p}}_m^T{{\bf{a}}^{(t)}},{\sigma ^2}\left\| {{{\bf{p}}_m}} \right\|_2^2} \right),
%\end{equation}
%respectively, since ${\bf{p}}_m^T{\bf{H}} = {\bf{0}}$. 

As a result, by replacing $\Lambda _{k,t}^{(K)}$ with $\tLambda _{k,t}^{(K)}$ in (\ref{GCUSUM_test}), a relaxed generalized CUSUM (RGCUSUM) test can be expressed as
\begin{equation} \label{GCUSUM_simplified}
{T_R} = \min \left\{ {K:\mathop {\max }\limits_{1 \le k \le K} \mathop {\sup }\limits_{\{ {{\cA}^{(t)}}\} } \sum\limits_{t = k}^K {\tLambda _{k,t}^{(K)}}  \ge h} \right\} = \min \left\{ {K:\mathop {\max }\limits_{1 \le k \le K} \mathop {\sup }\limits_{\{ {\cA^{(t)}}\} } \sum\limits_{t = k}^K {\sum\limits_{m \in {\cA^{(t)}}} {\zeta _m^{(t)}} }  \ge h} \right\}.
\end{equation}
It is worth mentioning that since
%\begin{equation}
%\mathop {\max }\limits_{1 \le k \le K} \mathop {\sup }\limits_{\{ {\cA^{(t)}}\} } \sum\limits_{t = k}^K {\sum\limits_{m \in {\cA^{(t)}}} {\zeta _m^{(t)}} }  = \mathop {\max }\limits_{1 \le k \le K} \mathop {\sup }\limits_{\{ {\cA^{(t)}}\} } \sum\limits_{t = k}^K { \tLambda _{k,t}^{(K)}}  \ge \mathop {\max }\limits_{1 \le k \le K} \mathop {\sup }\limits_{\{ {\cA^{(t)}}\} } \sum\limits_{t = k}^K {\Lambda _{k,t}^{(K)}}  = \mathop {\max }\limits_{1 \le k \le K} \mathop {\sup }\limits_{\{ {\cA^{(t)}}\} } \Lambda _k^{(K)}
%\end{equation}
\begin{equation}
\mathop {\max }\limits_{1 \le k \le K} \mathop {\sup }\limits_{\{ {\cA^{(t)}}\} } \sum\limits_{t = k}^K { \tLambda _{k,t}^{(K)}}  \ge \mathop {\max }\limits_{1 \le k \le K} \mathop {\sup }\limits_{\{ {\cA^{(t)}}\} } \sum\limits_{t = k}^K {\Lambda _{k,t}^{(K)}}  = \mathop {\max }\limits_{1 \le k \le K} \mathop {\sup }\limits_{\{ {\cA^{(t)}}\} } \Lambda _k^{(K)},
\end{equation}
we know that $T_R \le T_G$ for any given $h$, which implies that for a given $h$, the RGCUSUM test rule in (\ref{GCUSUM_simplified}) reduces the detection delay when compared to the GCUSUM test rule in (\ref{GCUSUM_test}) at the expense of a shorter false alarm period. 
%It is worth mentioning that this is sometimes more favorable for a monitoring system

%\cbr Add discussion about the impact on detection power and false alarm. \cbk
%
%\cbr(The reasons to employ $\tLambda _{k,t}^{(K)}$ are twofold. On one hand, there is no analytical expression for $\Lambda _{k,t}^{(K)}$. On the other hand, avoiding a miss detection is usually more favorable than avoiding a false alarm for a monitoring system.) \cbk

Let ${\omega ^{(K)}}$ denote the test statistic of the RGCUSUM at time instant $K$, which can be simplified to
\begin{align} \notag
{\omega ^{(K)}} & \buildrel \Delta \over 
%= \mathop {\max }\limits_{1 \le k \le K} \mathop {\sup }\limits_{\left\{ {{{\cal A}^{(t)}}} \right\}} \Lambda _k^K 
= \mathop {\max }\limits_{1 \le k \le K} \mathop {\sup }\limits_{\left\{ {{{\cal A}^{(t)}}} \right\}} \sum\limits_{t = k}^K {\sum\limits_{m \in {{\cal A}^{(t)}}} {\zeta _m^{(t)}} } \\ \notag
& = \mathop {\max }\limits_{1 \le k \le K} \sum\limits_{t = k}^K {\mathop {\sup }\limits_{ {{{\cal A}^{(t)}}} } \sum\limits_{m \in {{\cal A}^{(t)}}} {\zeta _m^{(t)}} } \\ \label{define_omega_K}
&  = \mathop {\max }\limits_{1 \le k \le K} \sum\limits_{t = k}^K {\sum\limits_{m = 1}^M {\max \left\{ {\zeta _m^{(t)},0} \right\}} }.
\end{align}
Noting that ${\max \{ {\zeta _m^{(t)},0} \}} \ge 0$, we can obtain
\begin{align} \notag
{\omega ^{(K)}} & = \mathop {\max }\limits_{1 \le k \le K} \sum\limits_{t = k}^K {\sum\limits_{m = 1}^M {\max \left\{ {\zeta _m^{(t)},0} \right\}} } \\ \notag
& = \max \left\{ {\mathop {\max }\limits_{1 \le k \le K - 1} \sum\limits_{t = k}^K {\sum\limits_{m = 1}^M {\max \left\{ {\zeta _m^{(t)},0} \right\}} } ,\sum\limits_{m = 1}^M {\max \left\{ {\zeta _m^{(K)},0} \right\}} } \right\}\\ \notag
%& = \max \left\{ {\mathop {\max }\limits_{1 \le k \le K - 1} \sum\limits_{t = k}^{K - 1} {\max \left\{ {\zeta _m^{(t)},0} \right\}} ,0} \right\} + \sum\limits_{m = 1}^M {\max \left\{ {\zeta _m^{(t)},0} \right\}} \\
& = \max \left\{ {{\omega ^{(K - 1)}},0} \right\} + \sum\limits_{m = 1}^M {\max \left\{ {\zeta _m^{(K)},0} \right\}} \\ \label{RGCUSUM_recursive}
& = {\omega ^{(K - 1)}} + \sum\limits_{m = 1}^M { {\max \left\{ {\zeta _m^{(K)},0} \right\}}} 
\end{align}
with $\omega ^{(0)} \triangleq 0$. Hence, $\omega ^{(K)}$ can be calculated recursively. Furthermore, from (\ref{RGCUSUM_recursive}), the test statistic of the RGCUSUM can be rewritten as
%\begin{align}
%\mathop {\max }\limits_{1 \le k \le K} \mathop {\sup }\limits_{\cal A} \Lambda _{k}^{(K)} = \sum\limits_{m = 1}^M {\sum\limits_{t = 1}^K {\xi _m^{(t)}}} 
%\end{align}
\begin{equation} \label{relaxed_GCUSUM}
%{\omega ^{(K)}}   = \sum\limits_{t = 1}^K {\sum\limits_{m= 1}^M {\tzeta _m^{(t)}} }.
{\omega ^{(K)}}   = \sum\limits_{t = 1}^K {\sum\limits_{m= 1}^M {{\max \left\{ {\zeta _m^{(t)},0} \right\}} } },
\end{equation}
and therefore, by employing (\ref{GCUSUM_simplified}), (\ref{define_omega_K}) and (\ref{relaxed_GCUSUM}), the stopping time of the RGCUSUM can be simplified to 
\begin{equation} \label{stopping_time_simplified}
{T_R} = \min \left\{ {K:{\omega ^{(K)}} \ge h} \right\} = \min \left\{ {K:\sum\limits_{t = 1}^K {\sum\limits_{m = 1}^M {\max \left\{ {\zeta _m^{(t)},0} \right\}} }  \ge h} \right\}.
\end{equation}
 The procedure for implementing the proposed RGCUSUM is summarized in Algorithm \ref{Algorithm_RGCUSUM}.  It is seen from (\ref{recursive_GCUSUM}) and (\ref{RGCUSUM_recursive}) that the test statistics of the GCUSUM and the RGCUSUM can be both expressed in a recursive way, which can greatly facilitate the computation of the corresponding test statistics in the sequential setting. However, it is worth pointing out that the computational complexity of the test statistic of the RGCUSUM is just linear in the number $M$ of observations since there is a closed-form expression for $\zeta _m^{(t)}$; while the computational complexity of the test statistic of the GCUSUM is exponential in $M$ as explained in Section \ref{Section_GCUSUM}. To this end, the RGCUSUM test rule is more amenable to implementation in practice, especially when $M$ is large.

\begin{algorithm}[htb]
	\caption{RGCUSUM Algorithm}
	\begin{algorithmic}[1]
		\STATE \textbf{Input}:  ${\bf x}^{(t)}$, $\bf H$, $\rho_L$, $\rho_U$, $\sigma^2$ and $h$
		\STATE  \textbf{Output}: $T_{{R}}$
		\STATE \textbf{Initialization}: $K \leftarrow 0$, $\omega^{(0)} \leftarrow 0$
		\WHILE{$\omega^{(K)} <h$}
		\STATE $K \leftarrow K+1$
		\STATE Evaluate $\zeta _m^{(K)}$ for all $m$ by using (\ref{tilde_zeta_m_t_define}). 
		\STATE Update the decision statistic: $\omega^{(K)} \leftarrow \omega^{(K-1)} + \sum_{m = 1}^M { {\max \{ {\zeta _m^{(K)},0} \}}}$.
%		\STATE Update the decision statistic: ${V_K} \leftarrow \max \left\{ {{V_{K - 1}},0} \right\} + {v_K}$.
		\ENDWHILE
		\STATE $T_{{R}}\leftarrow K$, declare an FDIA.
	\end{algorithmic}
	\label{Algorithm_RGCUSUM}
\end{algorithm}

% \cbr add discussion on the computational complexity of $\omega ^{(K)}$ when compared to GCUSUM \cbk

\section{Performance Analysis of the Proposed Relaxed Generalized CUSUM Test}
\label{Section_Performance_Analysis}

%\cbr Give the lemmas on the expectation of the increase \cbk

%\cbr Add discussion on $\zeta _m^{(t)}$: Under no attack, $\zeta _m^{(t)}<0$, while under attack, $\zeta _m^{(t)} \ge0$. \cbk

In this section, we provide the performance analysis of our proposed RGCUSUM test. In particular, we provide a sufficient condition under which the constraint on the expected false alarm period  in (\ref{changepoint_detection_problem_formulation}) can be guaranteed, which offers insight on the design of the proposed RGCUSUM test rule to achieve the prescribed performance requirement. Moreover, an upper bound on the worst-case expected detection delay defined in (\ref{lorden_formulation}) is also derived.

%${\zeta _m^{(t)},0}$ depends on ${\tilde{x}}_m^{(t)}$, so we first exam the property of ${\tilde{x}}_m^{(t)}$.

It is seen from (\ref{relaxed_GCUSUM}) that for a given $K$, the test statistic of the RGCUSUM is fully determined by the sequences $\{ {\max \{ {\zeta _m^{(t)},0} \}} \}_{t\ge1}$ for all $m$.
%, and moreover, $\zeta _m^{(t)}$ is just a function of $\tilde x_m^{(t)}$. 
In light of this, in order to investigate the performance of the RGCUSUM, we first characterize the statistical properties of the sequences $\{ {\max \{ {\zeta _m^{(t)},0} \}} \}_{t\ge1}$ for all $m$.

Let ${\bf p}_i^T$ denote the $i$-th row of the projection matrix ${\bf{P}}_{\bf{H}}^ \bot$, i.e.,
\begin{equation} \label{P_pm}
{\bf{P}}_{\bf{H}}^ \bot  = \left[ {\bf{p}}_1, {\bf{p}}_2, ..., {\bf{p}}_M  \right]^T
\end{equation}
%\begin{equation} \label{P_pm}
%{\bf{P}}_{\bf{H}}^ \bot  = \left( \begin{array}{l}
%{\bf{p}}_1^T\\
%{\bf{p}}_2^T\\
%\vdots \\
%{\bf{p}}_M^T
%\end{array} \right).
%\end{equation}
By employing the definition of ${\tilde {\bf x}^{(t)}}$ in (\ref{tilde_x_define}), we can obtain
\begin{equation}
\tilde x_m^{(t)} = {\bf{p}}_m^T{{\bf{x}}^{(t)}},
\end{equation}
which together with (\ref{problem_formulation}) yields that the pre- and post-attack distributions of $\tilde x_m^{(t)}$ can be expressed as
%\begin{equation} \label{tilde_x_distribution}
%\tilde x_m^{(t)} \sim {\cal N}\left( {0,{\sigma ^2}\left\| {{{\bf{p}}_m}} \right\|_2^2} \right) \quad \text{ and } \quad \tilde x_m^{(t)} \sim {\cal N}\left( {{\bf{p}}_m^T{{\bf{a}}^{(t)}},{\sigma ^2}\left\| {{{\bf{p}}_m}} \right\|_2^2} \right),
%\end{equation}
\begin{equation} \label{tilde_x_distribution_0}
\tilde x_m^{(t)} \sim {\cal N}\left( {0,{\sigma ^2}\left\| {{{\bf{p}}_m}} \right\|_2^2} \right)
\end{equation}
\begin{equation} \label{tilde_x_distribution_1}
\text{and} \quad \tilde x_m^{(t)} \sim {\cal N}\left( {{\bf{p}}_m^T{{\bf{a}}^{(t)}},{\sigma ^2}\left\| {{{\bf{p}}_m}} \right\|_2^2} \right),
\end{equation}
respectively, since ${\bf{p}}_m^T{\bf{H}} = {\bf{0}}$ due to the definition of ${\bf{P}}_{\bf{H}}^ \bot$ in (\ref{Define_P_H_per}). 

It is seen from (\ref{problem_formulation}) that since ${\btheta}^{(t)}$ is time-varying over $t$, both the pre-attack observation sequence $\{{\bf x}^{(t)}\}_{t< t_a}$ and the post-attack observation sequence $\{{\bf x}^{(t)}\}_{t\ge t_a}$ can be non-identically distributed. In addition, for two distinct attack time instants $t_a$ and $t_a'$, the corresponding observations ${\bf x}^{(t_a +t)}$ and ${\bf x}^{(t_a' +t)}$ are not necessarily identically distributed for any $t \ge 0$. However, it is seen from (\ref{tilde_x_distribution_0}) that the pre-attack observation sequence $\{\tilde x_m^{(t)} \}_{t< t_a}$ is an i.i.d sequence. Moreover, as demonstrated by (\ref{tilde_x_distribution_1}), for two distinct attack time instants $t_a$ and $t_a'$, the corresponding observations ${\tilde x_m^{(t_a +t)}}$ and ${\tilde x_m^{(t_a' +t)}}$ are identically distributed for any $t \ge 0$, since ${\bf a}^{(t_a + t-1)} = {\bf b}^{(t)}$ for any $t_a$ and $t \ge 1$.

A particular note of interest is that if there is no attack and $\sigma^2$ is small, then $\tilde x_m^{(t)}$ is very likely to be close to $0$ from (\ref{tilde_x_distribution_0}), which gives rise to ${\max \{ {\zeta _m^{(t)},0} \}}=0$ since from (\ref{tilde_zeta_m_t_define}), $\zeta _m^{(t)}=\frac{1}{{2{\sigma ^2}}}( {2| {{\tilde x_m^{(t)}}} |\rho_L  - {\rho_L ^2}} )<0$ when $\tilde x_m^{(t)}$ is close to $0$. Hence, it is seen from (\ref{relaxed_GCUSUM}) that the test statistic ${\omega ^{(K)}} $ of the RGCUSUM is very likely to remain  close to $0$ as $K$ grows. On the contrary, if attacks occur at time instant $t_a$ and ${\bf{p}}_m^T{{\bf{a}}^{(t)}}$ is large for $t\ge t_a$, then from (\ref{tilde_x_distribution_1}), a large $\tilde x_m^{(t)}$ is more likely to be observed for $t\ge t_a$ which brings about $\zeta _m^{(t)} >0$ for $t\ge t_a$ from (\ref{tilde_zeta_m_t_define}), and hence ${\max \{ {\zeta _m^{(t)},0} \}}>0$ for $t\ge t_a$. Therefore, ${\omega ^{(K)}} $ increases fast as $K$ grows when $K \ge t_a$. To this end, the false alarm period and the detection delay of the RGCUSUM are expected to be large and small, respectively.

%Let's first look at the statistic ${\max \{ {\zeta _m^{(t)},0} \}}$.

In the following Lemma, we provide an upper bound and a lower bound on the expectation of the statistic $\{ {\max \{ {\zeta _m^{(t)},0} \}} \}$ with respect to the true distributions of ${\bf x}^{(1)}$, ${\bf x}^{(2)}$,  ${\bf x}^{(3)}$, ... when the attacks never occur or occur at time instant $t_a$, respectively.

\begin{lemma} \label{Lemma_bounds_on_expectations}
	\begin{equation} \label{expected_max_zeta_0_ub}
	{\bbE_{\infty}}\left\{ {\max \left\{ {\zeta _m^{(t)},0} \right\}} \right\} \le \frac{1}{2}\left\| {{{\bf{p}}_m}} \right\|_2^2 + \frac{{{\rho _L} + {\rho _U}}}{\sigma }{\left\| {{{\bf{p}}_m}} \right\|_2}\sqrt {\frac{2}{\pi }}, 
	\end{equation}
	\begin{equation} \label{expected_max_zeta_1_lb}
	{\bbE_{t_a}}\left\{ {\max \left\{ {\zeta _m^{(t)},0} \right\}} \right\} \ge \frac{{\rho _L^2}}{{2{\sigma ^2}}}\left\{ {{\rm{erf}}\left( {\frac{{2{\rho _U}}}{{\sqrt 2 \sigma {{\left\| {{{\bf{p}}_m}} \right\|}_2}}}} \right) - {\rm{erf}}\left( {\frac{{{\rho _L} + {\rho _U}}}{{\sqrt 2 \sigma {{\left\| {{{\bf{p}}_m}} \right\|}_2}}}} \right)} \right\}, \quad \forall t \ge t_a
	\end{equation}
	where the error function 
	\begin{equation} \label{error_function_define}
	{\rm{erf}}\left( x \right) \buildrel \Delta \over = \frac{2}{{\sqrt \pi  }}\int_0^x {{e^{ - {s^2}}}ds}.
	\end{equation}
\end{lemma}

\begin{IEEEproof}
Let's first prove (\ref{expected_max_zeta_0_ub}).
%consider the expectation of ${\max \{ {\zeta _m^{(t)},0} \}}$ with respect to the pre-attack probability measure. 
Notice from (\ref{tilde_zeta_m_t_define}) that
\begin{equation}
\left\{ \begin{array}{l}
\zeta _m^{(t)} = \frac{1}{{2{\sigma ^2}}}{\left| {{\tilde x_m^{(t)}}} \right|^2} \;\; \;\; \text{if} \;\; {\rho_L} \le \left| {{\tilde x_m^{(t)}}} \right| \le  {\rho_U},\\
\zeta _m^{(t)} \le \frac{{{\rho _L}}}{{{\sigma ^2}}}\left| {\tilde x_m^{(t)}} \right| \; \;\; \;\; \text{if} \;\; \left| {{\tilde x_m^{(t)}}} \right| < {\rho_L},\\
\zeta _m^{(t)} \le \frac{{{\rho _U}}}{{{\sigma ^2}}}\left| {\tilde x_m^{(t)}} \right| \; \;\; \;\; \text{if} \;\; \left| {{\tilde x_m^{(t)}}} \right| > {\rho_U},
\end{array} \right.
\end{equation}
which yields
\begin{align} \notag
& {\bbE_{\infty}}\left\{ {\max \left\{ {\zeta _m^{(t)},0} \right\}} \right\} \\  \notag
& = {\bbE_{\infty}}\left\{ \max \left\{ {\zeta _m^{(t)},0} \right\}{\bone\left\{ {\left| {\tilde x_m^{(t)}} \right| < {\rho _L}} \right\}} \right\} + {\bbE_{\infty}}\left\{\max \left\{ {\zeta _m^{(t)},0} \right\} {\bone\left\{ {{\rho _L} \le \left| {\tilde x_m^{(t)}} \right| \le {\rho _U}} \right\}} \right\} \\  \notag
& \quad + {\bbE_{\infty}}\left\{ \max \left\{ {\zeta _m^{(t)},0} \right\}{\bone\left\{ {\left| {\tilde x_m^{(t)}} \right| > {\rho _U}} \right\}} \right\}\\  \notag
& \le {\bbE_{\infty}}\left\{ \frac{{{\rho _L}}}{{{\sigma ^2}}}\left| {\tilde x_m^{(t)}} \right|{\bone\left\{ {\left| {\tilde x_m^{(t)}} \right| < {\rho _L}} \right\}} \right\} + {\bbE_{\infty}}\left\{ \frac{1}{{2{\sigma ^2}}}{{\left| {\tilde x_m^{(t)}} \right|}^2}{\bone\left\{ {{\rho _L} \le \left| {\tilde x_m^{(t)}} \right| \le {\rho _U}} \right\}} \right\} \\    \notag
& \quad + {\bbE_{\infty}}\left\{ \frac{{{\rho _U}}}{{{\sigma ^2}}}\left| {\tilde x_m^{(t)}} \right|{\bone\left\{ {\left| {\tilde x_m^{(t)}} \right| > {\rho _U}} \right\}} \right\} \\ \label{E_0_max_zeta_ub1}
& \le {\bbE_{\infty}}\left\{ {\frac{{{\rho _L} + {\rho _U}}}{{{\sigma ^2}}}\left| {\tilde x_m^{(t)}} \right|} \right\} + {\bbE_{\infty}}\left\{ {\frac{1}{{2{\sigma ^2}}}{{\left| {\tilde x_m^{(t)}} \right|}^2}} \right\}.
\end{align}

If there is no attack, then by employing (\ref{tilde_x_distribution_0}), we know that ${| {\tilde x_m^{(t)}} |}$ and ${{| {\tilde x_m^{(t)}} |^2}}$ follow a folded normal distribution \cite{leone1961folded} and a chi-squared distribution, respectively, and therefore, we can obtain
%\begin{equation}
%{\bbE_{\infty}}\left\{ {\left| {\tilde x_m^{(t)}} \right|} \right\} = \sigma {\left\| {{{\bf{p}}_m}} \right\|_2}\sqrt {\frac{2}{\pi }} \quad \text{ and } \quad {\bbE_{\infty}}\left\{ {{{\left| {\tilde x_m^{(t)}} \right|}^2}} \right\} = {\sigma ^2}\left\| {{{\bf{p}}_m}} \right\|_2^2,
%\end{equation}
\begin{equation}
{\bbE_{\infty}}\left\{ {\left| {\tilde x_m^{(t)}} \right|} \right\} = \sigma {\left\| {{{\bf{p}}_m}} \right\|_2}\sqrt {\frac{2}{\pi }} \quad \text{and} \quad {\bbE_{\infty}}\left\{ {{{\left| {\tilde x_m^{(t)}} \right|}^2}} \right\} = {\sigma ^2}\left\| {{{\bf{p}}_m}} \right\|_2^2,
\end{equation}
%\begin{equation}
%
%\end{equation}
which implies
\begin{equation}
{\bbE_{\infty}}\left\{ {\max \left\{ {\zeta _m^{(t)},0} \right\}} \right\} \le \frac{1}{2}\left\| {{{\bf{p}}_m}} \right\|_2^2 + \frac{{{\rho _L} + {\rho _U}}}{\sigma }{\left\| {{{\bf{p}}_m}} \right\|_2}\sqrt {\frac{2}{\pi }}
\end{equation}
by employing (\ref{E_0_max_zeta_ub1}), which completes the proof for (\ref{expected_max_zeta_0_ub}).

Now, we prove (\ref{expected_max_zeta_1_lb}). Since ${\max \{ {\zeta _m^{(t)},0} \}}\ge0$, by employing (\ref{tilde_zeta_m_t_define}), we can obtain
\begin{align} \notag
& {\bbE_{t_a}}\left\{ {\max \left\{ {\zeta _m^{(t)},0} \right\}} \right\} \\  \notag
& = {\bbE_{t_a}}\left\{ \max \left\{ {\zeta _m^{(t)},0} \right\}{\bone\left\{ {\left| {\tilde x_m^{(t)}} \right| < {\rho _L}} \right\}} \right\} + {\bbE_{t_a}}\left\{\max \left\{ {\zeta _m^{(t)},0} \right\} {\bone\left\{ {{\rho _L} \le \left| {\tilde x_m^{(t)}} \right| \le {\rho _U}} \right\}} \right\} \\ \notag
& \quad + {\bbE_{t_a}}\left\{ \max \left\{ {\zeta _m^{(t)},0} \right\}{\bone\left\{ {\left| {\tilde x_m^{(t)}} \right| > {\rho _U}} \right\}} \right\}\\  \notag
&\ge {\bbE_{t_a}}\left\{ {\frac{1}{{2{\sigma ^2}}}{\left| {\tilde x_m^{(t)}} \right|^2}\bone\left\{ {{\rho _L} \le \left| {\tilde x_m^{(t)}} \right| \le {\rho _U}} \right\}} \right\}\\ \notag
&\ge \frac{{\rho _L^2}}{{2{\sigma ^2}}}{\bbE_{t_a}}\left\{ {\bone\left\{ {{\rho _L} \le \left| {\tilde x_m^{(t)}} \right| \le {\rho _U}} \right\}} \right\}\\ \label{expected_max_zeta_0_lb_1}
&= \frac{{\rho _L^2}}{{2{\sigma ^2}}}{\bbP_{t_a}}\left( {{\rho _L} \le \left| {\tilde x_m^{(t)}} \right| \le {\rho _U}} \right), \quad \forall t \ge t-a,
\end{align}
where ${\bbP_{t_a}}()$ is the probability measure when the attack time is $t_a$.

Noting that 
%in the presence of attacks, 
if $t\ge t_a$, then by employing (\ref{tilde_x_distribution_1}), the probability density function of $| {\tilde x_m^{(t)}} |$ can be written as
\begin{equation}
\frac{1}{{\sqrt {2\pi } \sigma {{\left\| {{{\bf{p}}_m}} \right\|}_2}}}\Bigg[ {{e^{ - \frac{{{{\left( {\left| {\tilde x_m^{(t)}} \right| - {\bf{p}}_m^T{{\bf{a}}^{(t)}}} \right)}^2}}}{{2{\sigma ^2}\left\| {{{\bf{p}}_m}} \right\|_2^2}}}} + {e^{ - \frac{{{{\left( {\left| {\tilde x_m^{(t)}} \right| + {\bf{p}}_m^T{{\bf{a}}^{(t)}}} \right)}^2}}}{{2{\sigma ^2}\left\| {{{\bf{p}}_m}} \right\|_2^2}}}}} \Bigg],
\end{equation}
which yields
\begin{align} \notag
{\bbP_{t_a}}\left( {{\rho _L} \le \left| {\tilde x_m^{(t)}} \right| \le {\rho _U}} \right) & = \frac{1}{2}\left[ {{\rm{erf}}\left( {\frac{{{\rho _U} + {\bf{p}}_m^T{{\bf{a}}^{(t)}}}}{{\sqrt 2 \sigma {{\left\| {{{\bf{p}}_m}} \right\|}_2}}}} \right) + {\rm{erf}}\left( {\frac{{{\rho _U} - {\bf{p}}_m^T{{\bf{a}}^{(t)}}}}{{\sqrt 2 \sigma {{\left\| {{{\bf{p}}_m}} \right\|}_2}}}} \right)} \right] \\  \notag
& \quad - \frac{1}{2}\left[ {{\rm{erf}}\left( {\frac{{{\rho _L} + {\bf{p}}_m^T{{\bf{a}}^{(t)}}}}{{\sqrt 2 \sigma {{\left\| {{{\bf{p}}_m}} \right\|}_2}}}} \right) + {\rm{erf}}\left( {\frac{{{\rho _L} - {\bf{p}}_m^T{{\bf{a}}^{(t)}}}}{{\sqrt 2 \sigma {{\left\| {{{\bf{p}}_m}} \right\|}_2}}}} \right)} \right]\\  \notag
&= \frac{1}{2}\underbrace {\left[ {{\rm{erf}}\left( {\frac{{{\rho _U} + \mu _m^{(t)}}}{{\sqrt 2 \sigma {{\left\| {{{\bf{p}}_m}} \right\|}_2}}}} \right) - {\rm{erf}}\left( {\frac{{{\rho _L} + \mu _m^{(t)}}}{{\sqrt 2 \sigma {{\left\| {{{\bf{p}}_m}} \right\|}_2}}}} \right)} \right]}_{ \buildrel \Delta \over = {g_1}\left( {\mu _m^{(t)}} \right)} \\  \label{P_1_x_m_t}
& \quad + \frac{1}{2}\underbrace {\left[ {{\rm{erf}}\left( {\frac{{{\rho _U} - \mu _m^{(t)}}}{{\sqrt 2 \sigma {{\left\| {{{\bf{p}}_m}} \right\|}_2}}}} \right) - {\rm{erf}}\left( {\frac{{{\rho _L} - \mu _m^{(t)}}}{{\sqrt 2 \sigma {{\left\| {{{\bf{p}}_m}} \right\|}_2}}}} \right)} \right]}_{ \buildrel \Delta \over = {g_2}\left( {\mu _m^{(t)}} \right)}, \quad \forall t\ge t_a,
\end{align}
where (\ref{P_1_x_m_t}) is due to the fact that $\mu _m^{(t)} = {\bf{p}}_m^T{{\bf{a}}^{(t)}}$ by employing (\ref{Define_mu}) and (\ref{P_pm}).

By using (\ref{error_function_define}), ${g_1}( {\mu _m^{(t)}} )$ can be rewritten as
\begin{equation}
{g_1}\left( {\mu _m^{(t)}} \right) = \frac{2}{{\sqrt \pi  }}\int_{\frac{{{\rho _L} + \mu _m^{(t)}}}{{\sqrt 2 \sigma {{\left\| {{{\bf{p}}_m}} \right\|}_2}}}}^{\frac{{{\rho _U} + \mu _m^{(t)}}}{{\sqrt 2 \sigma {{\left\| {{{\bf{p}}_m}} \right\|}_2}}}} {{e^{ - {s^2}}}ds}
\end{equation}
which is neither convex nor concave. By taking the derivative of ${g_1}( {\mu _m^{(t)}} ) $ with respect to ${\mu _m^{(t)}}$, we can obtain
\begin{equation} \label{derivative_of_g_1}
\frac{{d{g_1}\left( {\mu _m^{(t)}} \right)}}{{d\mu _m^{(t)}}} = \frac{2}{{\sqrt {2\pi } \sigma {{\left\| {{{\bf{p}}_m}} \right\|}_2}}}\Bigg[ {{e^{ - \frac{{{{\left( {{\rho _U} + \mu _m^{(t)}} \right)}^2}}}{{2{\sigma ^2}\left\| {{{\bf{p}}_m}} \right\|_2^2}}}} - {e^{ - \frac{{{{\left( {{\rho _L} + \mu _m^{(t)}} \right)}^2}}}{{2{\sigma ^2}\left\| {{{\bf{p}}_m}} \right\|_2^2}}}}} \Bigg].
\end{equation}
It is seen from (\ref{derivative_of_g_1}) that
\begin{equation}
\left\{ \begin{array}{l}
\frac{{d{g_1}\left( {\mu _m^{(t)}} \right)}}{{d\mu _m^{(t)}}} \ge 0, \quad \text{if} \quad \mu _m^{(t)} \le  - \frac{{{\rho _L} + {\rho _U}}}{2},\\
\frac{{d{g_1}\left( {\mu _m^{(t)}} \right)}}{{d\mu _m^{(t)}}} < 0, \quad \text{if} \quad \mu _m^{(t)} >  - \frac{{{\rho _L} + {\rho _U}}}{2}.
\end{array} \right.
\end{equation}
which implies that ${g_1}( {\mu _m^{(t)}} )$ achieves its minimum at either ${\mu _m^{(t)}} = - \rho_U$ or ${\mu _m^{(t)}} = \rho_U$, since  $ -\rho_U\le {\mu _m^{(t)}} \le \rho_U$ due to the constraint in (\ref{mu_constraint}). Moreover, by employing (\ref{derivative_of_g_1}), we know that for any $s$,
\begin{equation}
{\left. {\frac{{d{g_1}\left( {\mu _m^{(t)}} \right)}}{{d\mu _m^{(t)}}}} \right|_{ - \frac{{{\rho _L} + {\rho _U}}}{2} + s}} = \frac{2}{{\sqrt {2\pi } \sigma {{\left\| {{{\bf{p}}_m}} \right\|}_2}}}\Bigg[ {{e^{ - \frac{{{{\left( {\frac{{{\rho _U} - {\rho _L}}}{2} + s} \right)}^2}}}{{2{\sigma ^2}\left\| {{{\bf{p}}_m}} \right\|_2^2}}}} - {e^{ - \frac{{{{\left( {\frac{{{\rho _U} - {\rho _L}}}{2} - s} \right)}^2}}}{{2{\sigma ^2}\left\| {{{\bf{p}}_m}} \right\|_2^2}}}}} \Bigg] =  - {\left. {\frac{{d{g_1}\left( {\mu _m^{(t)}} \right)}}{{d\mu _m^{(t)}}}} \right|_{ - \frac{{{\rho _L} + {\rho _U}}}{2} - s}},
\end{equation}
and hence $\frac{{d{g_1}\left( {\mu _m^{(t)}} \right)}}{{d\mu _m^{(t)}}}$ is an odd function with respect to ${\mu _m^{(t)}} = -\frac{\rho_L + \rho_U}{2}$. Therefore, ${g_1}( {\mu _m^{(t)}} )$ achieves its minimum at ${\mu _m^{(t)}} = \rho_U$, that is,
\begin{equation} \label{g_1_lb}
{g_1}\left( {\mu _m^{(t)}} \right) \ge {g_1}\left( {{\rho _U}} \right) = {\rm{erf}}\left( {\frac{{2{\rho _U}}}{{\sqrt 2 \sigma {{\left\| {{{\bf{p}}_m}} \right\|}_2}}}} \right) - {\rm{erf}}\left( {\frac{{{\rho _L} + {\rho _U}}}{{\sqrt 2 \sigma {{\left\| {{{\bf{p}}_m}} \right\|}_2}}}} \right).
\end{equation}
%since $0 \le {\rho _L} \le {\rho _U}$ and $-{\rho _U} \le {\mu _m^{(t)}} \le {\rho _U}$ according to (\ref{mu_constraint}). 
Similarly, we can show that
\begin{equation} \label{g_2_lb}
{g_2}\left( {\mu _m^{(t)}} \right) \ge {g_2}\left( -{{\rho _U}} \right) = {\rm{erf}}\left( {\frac{{2{\rho _U}}}{{\sqrt 2 \sigma {{\left\| {{{\bf{p}}_m}} \right\|}_2}}}} \right) - {\rm{erf}}\left( {\frac{{{\rho _L} + {\rho _U}}}{{\sqrt 2 \sigma {{\left\| {{{\bf{p}}_m}} \right\|}_2}}}} \right).
\end{equation}
By employing (\ref{expected_max_zeta_0_lb_1}), (\ref{P_1_x_m_t}), (\ref{g_1_lb}) and (\ref{g_2_lb}), we can obtain
\begin{equation}
{\bbE_{t_a}}\left\{ {\max \left\{ {\zeta _m^{(t)},0} \right\}} \right\} \ge \frac{{\rho _L^2}}{{2{\sigma ^2}}}\left\{ {{\rm{erf}}\left( {\frac{{2{\rho _U}}}{{\sqrt 2 \sigma {{\left\| {{{\bf{p}}_m}} \right\|}_2}}}} \right) - {\rm{erf}}\left( {\frac{{{\rho _L} + {\rho _U}}}{{\sqrt 2 \sigma {{\left\| {{{\bf{p}}_m}} \right\|}_2}}}} \right)} \right\}, \quad \forall t\ge t_a,
\end{equation}
which completes the proof.
\end{IEEEproof}

\subsection{Conditions to Meet Expected False Alarm Period Constraint}

Motivated by the FDIA detection problem formulated in (\ref{changepoint_detection_problem_formulation}), the expected running length of the RGCUSUM under no attack needs to be guaranteed to be larger than the required lower bound $\gamma$ to avoid frequent false alarms. In general, we can set the threshold $h$ in (\ref{GCUSUM_simplified}) to be sufficiently large so that the constraint in (\ref{changepoint_detection_problem_formulation}) is satisfied. In the following, we provide a sufficient condition on $h$ which can guarantee that the expected false alarm period of the RGCUSUM is larger than the prescribed $\gamma$.

\begin{theorem} \label{Theorem_false_alarm_period}
	The constraint on the expected false alarm period in (\ref{changepoint_detection_problem_formulation}) is satisfied provided that
	\begin{equation} \label{h_lower_bound}
	h \ge \gamma \sum\limits_{m = 1}^M {\left( {\frac{1}{2}\left\| {{{\bf{p}}_m}} \right\|_2^2 + \frac{{{\rho _L} + {\rho _U}}}{\sigma }{{\left\| {{{\bf{p}}_m}} \right\|}_2}\sqrt {\frac{2}{\pi }} } \right)}.
	\end{equation}
\end{theorem}

\begin{IEEEproof}
By employing (\ref{relaxed_GCUSUM}), we can obtain
\begin{equation}
{\bbE_{\infty}}\left\{ {{\omega ^{(T_R)}}} \right\} = {\bbE_{\infty}}\left\{ {\sum\limits_{t = 1}^{T_R} {\sum\limits_{m = 1}^M {\max \left\{ {\zeta _m^{(t)},0} \right\}} } } \right\} = {\bbE_{\infty}}\left\{ {\sum\limits_{t = 1}^\infty  {\sum\limits_{m = 1}^M {\max \left\{ {\zeta _m^{(t)},0} \right\}\bone\left\{ {T_R \ge t} \right\}} } } \right\}.
\end{equation}
Noting that ${\max \{ {\zeta _m^{(t)},0} \}}$ and $\bone\left\{ {T_R \ge t} \right\}$ are independent, and ${\max \{ {\zeta _m^{(t)},0} \}}\ge0$, by employing the monotone convergence theorem, ${\bbE_{\infty}}\{ {{\omega ^{(K)}}} \}$ can be simplified to
\begin{align} \notag
{\bbE_{\infty}}\left\{ {{\omega ^{(T_R)}}} \right\} & = \sum\limits_{t = 1}^\infty  {{\bbE_{\infty}}\left\{ {\sum\limits_{m = 1}^M {\max \left\{ {\zeta _m^{(t)},0} \right\}\bone\left\{ {T_R \ge t} \right\}} } \right\}} \\ \notag
& = \sum\limits_{t = 1}^\infty  {\sum\limits_{m = 1}^M {{\bbE_{\infty}}\left\{ {\max \left\{ {\zeta _m^{(t)},0} \right\}} \right\}{\bbE_{\infty}}\left\{ {\bone\left\{ {T_R \ge t} \right\}} \right\}} } \\ \label{proof_theorem_false_alarm_1}
& = \sum\limits_{t = 1}^\infty  {\sum\limits_{m = 1}^M {{\bbE_{\infty}}\left\{ {\max \left\{ {\zeta _m^{(t)},0} \right\}} \right\}{\bbP_{\infty}}\left( {T_R \ge t} \right)} },
\end{align}
where $\bbP_{\infty}$ is the probability measure under no attack.

From (\ref{expected_max_zeta_0_ub}) in Lemma \ref{Lemma_bounds_on_expectations} and (\ref{proof_theorem_false_alarm_1}), ${\bbE_{\infty}}\{ {{\omega ^{(T_R)}}} \}$ can be bounded from above as per
\begin{align} \notag
{\bbE_{\infty}}\left\{ {{\omega ^{(T_R)}}} \right\} & \le \sum\limits_{t = 1}^\infty  {\sum\limits_{m = 1}^M {\left( {\frac{1}{2}\left\| {{{\bf{p}}_m}} \right\|_2^2 + \frac{{{\rho _L} + {\rho _U}}}{\sigma }{{\left\| {{{\bf{p}}_m}} \right\|}_2}\sqrt {\frac{2}{\pi }} } \right){\bbP_{\infty}}\left( {T_R \ge t} \right)} } \\ \notag
& = \sum\limits_{m = 1}^M {\left( {\frac{1}{2}\left\| {{{\bf{p}}_m}} \right\|_2^2 + \frac{{{\rho _L} + {\rho _U}}}{\sigma }{{\left\| {{{\bf{p}}_m}} \right\|}_2}\sqrt {\frac{2}{\pi }} } \right)} \sum\limits_{t = 1}^\infty  {{\bbP_{\infty}}\left( {T_R \ge t} \right)} \\ \label{proof_theorem_false_alarm_2}
& = \sum\limits_{m = 1}^M {\left( {\frac{1}{2}\left\| {{{\bf{p}}_m}} \right\|_2^2 + \frac{{{\rho _L} + {\rho _U}}}{\sigma }{{\left\| {{{\bf{p}}_m}} \right\|}_2}\sqrt {\frac{2}{\pi }} } \right)} {\bbE_{\infty}}\left\{ T_R \right\}.
\end{align}
Since when a false alarm is raised, ${\omega ^{(T_R)}} \ge h$ which implies
\begin{equation}
{\bbE_{\infty}}\left\{ {{\omega ^{(T_R)}}} \right\} \ge h,
\end{equation}
and therefore,
\begin{equation}
{\bbE_{\infty}}\left\{ {{T_R}} \right\} \ge \frac{h}{{\sum\limits_{m = 1}^M {\left( {\frac{1}{2}\left\| {{{\bf{p}}_m}} \right\|_2^2 + \frac{{{\rho _L} + {\rho _U}}}{\sigma }{{\left\| {{{\bf{p}}_m}} \right\|}_2}\sqrt {\frac{2}{\pi }} } \right)} }}
\end{equation}
by employing (\ref{proof_theorem_false_alarm_2}). Thus, if
\begin{equation}
h \ge \gamma \sum\limits_{m = 1}^M {\left( {\frac{1}{2}\left\| {{{\bf{p}}_m}} \right\|_2^2 + \frac{{{\rho _L} + {\rho _U}}}{\sigma }{{\left\| {{{\bf{p}}_m}} \right\|}_2}\sqrt {\frac{2}{\pi }} } \right)},
\end{equation}
then ${\bbE_{\infty}}\{ {{T_R}} \} \ge \gamma$ is guaranteed, which completes the proof.
\end{IEEEproof}

As demonstrated by Theorem \ref{Theorem_false_alarm_period}, in order to meet the constraint on the expected false alarm period, the ratio of the threshold $h$ employed in the RGCUSUM to the required lower bound $\gamma$ on the expected false alarm period should be larger than some constant.  It is worth mentioning that this constant is only determined by the projection operator ${\bf P}_{\bf H}^\perp$, the variance of the noise, and the prescribed lower and upper bounds on  the magnitude of the nonzero elements of $\bmu^{(t)}$. Therefore, when $\gamma$ is given, the lower bound on $h$ in (\ref{h_lower_bound}) can be calculated beforehand, and then employed in the RGCUSUM. Hence, Theorem \ref{Theorem_false_alarm_period} provides a guideline for the design of the proposed RGCUSUM to achieve the prescribed performance requirement.

\subsection{Upper Bound on the Worst-Case Expected Detection Delay}

Besides the expected false alarm period, another key performance measure for quickest detection is the worst-case expected detection delay. In this subsection, we scrutinize the worst-case expected detection delay of the proposed RGCUSUM.

By plugging (\ref{GCUSUM_simplified}) into (\ref{lorden_formulation}),  the worst-case expected detection delay of the proposed RGCUSUM can be expressed as
\begin{equation} \label{lorden_formulation_TR}
J\left( T_R \right) \buildrel \Delta \over = \mathop {\sup }\limits_{t_a}  {J_{{t_a}}}\left( {{T_R}} \right),
%{\bbE_{{t_a}}}\left\{ {\left. {{{\left( {{T_R} - {t_a} + 1} \right)}^ + }} \right|{{\cal F}_{{t_a} - 1}}} \right\}.
\end{equation}
where
\begin{equation} \label{Define_J_t_a_TR}
{J_{{t_a}}}\left( {{T_R}} \right) \buildrel \Delta \over = {\esssup _{{{\cal F}_{{t_a} - 1}}}} {\bbE_{{t_a}}}\left\{ {\left. {{{\left( {{T_R} - {t_a} + 1} \right)}^ + }} \right|{{\cal F}_{{t_a} - 1}}} \right\}.
\end{equation}

In order to derive an upper bound on $J\left( T_R \right)$ in (\ref{lorden_formulation_TR}), we first provide a lemma which specifies an upper bound on ${\bbE_{{t_a}}}\{ { {{{( {{T_R} - {t_a} + 1} )}^ + }} |{{\cal F}_{{t_a} - 1}}} \}$.

\begin{lemma} \label{Lemma_Conditional_Expectation_ub}
	For any given $t_a$, we have
	\begin{equation} 
	{\bbE_{{t_a}}}\left\{ {\left. {{{\left( {{T_R} - {t_a} + 1} \right)}^ + }} \right|{{\cal F}_{{t_a} - 1}}} \right\} \le \frac{{{\bbE_{{t_a}}}\left\{ {\left. {{\omega ^{({T_R})}}\bone\left\{ {{T_R} \ge {t_a}} \right\}} \right|{{\cal F}_{{t_a} - 1}}} \right\}}}{{\sum\limits_{m = 1}^M {\frac{{\rho _L^2}}{{2{\sigma ^2}}}\left\{ {{\rm{erf}}\left( {\frac{{2{\rho _U}}}{{\sqrt 2 \sigma {{\left\| {{{\bf{p}}_m}} \right\|}_2}}}} \right) - {\rm{erf}}\left( {\frac{{{\rho _L} + {\rho _U}}}{{\sqrt 2 \sigma {{\left\| {{{\bf{p}}_m}} \right\|}_2}}}} \right)} \right\}} }}.
	\end{equation}
\end{lemma}

\begin{IEEEproof}
Before proceeding, we first claim that
\begin{align} \notag
& {\bbE_{{t_a}}}\left\{ \bone\left\{ {{T_R} \ge {t_a}} \right\}{\left. {\sum\limits_{t = {t_a}}^{{T_R}} {\sum\limits_{m = 1}^M {\max \left\{ {\zeta _m^{(t)},0} \right\}} } } \right|{{\cal F}_{{t_a-1}}}} \right\} \\  \label{claim_1}
& \ge {\bbE_{{t_a}}}\left\{ {\left. {{{\left( {{T_R} - {t_a} + 1} \right)}^ + }} \right|{{\cal F}_{{t_a-1}}}} \right\}\sum\limits_{m = 1}^M {\frac{{\rho _L^2}}{{2{\sigma ^2}}}\left\{ {{\rm{erf}}\left( {\frac{{2{\rho _U}}}{{\sqrt 2 \sigma {{\left\| {{{\bf{p}}_m}} \right\|}_2}}}} \right) - {\rm{erf}}\left( {\frac{{{\rho _L} + {\rho _U}}}{{\sqrt 2 \sigma {{\left\| {{{\bf{p}}_m}} \right\|}_2}}}} \right)} \right\}}.
\end{align}
This claim can be proved as follows.
Note that
\begin{align}
& {\bbE_{{t_a}}}\left\{ \bone\left\{ {{T_R} \ge {t_a}} \right\}{\left. {\sum\limits_{t = {t_a}}^{{T_R}} {\sum\limits_{m = 1}^M {\max \left\{ {\zeta _m^{(t)},0} \right\}} } } \right|{{\cal F}_{{t_a-1}}}} \right\} \\ \label{explain_1_0}
& = \bone\left\{ {{T_R} \ge {t_a}} \right\}{\bbE_{{t_a}}}\left\{ {\left. {\sum\limits_{t = {t_a}}^\infty  {\sum\limits_{m = 1}^M {\max \left\{ {\zeta _m^{(t)},0} \right\}} \bone\left\{ {{T_R} \ge t} \right\}} } \right|{{\cal F}_{{t_a-1}}}} \right\}\\ \label{explain_1_1}
&= \bone\left\{ {{T_R} \ge {t_a}} \right\} \sum\limits_{t = {t_a}}^\infty  {{\bbE_{{t_a}}}\left\{ {\left. {\sum\limits_{m = 1}^M {\max \left\{ {\zeta _m^{(t)},0} \right\}} \bone\left\{ {{T_R} \ge t} \right\}} \right|{{\cal F}_{{t_a-1}}}} \right\}} \\ \label{explain_5_1}
& = \bone\left\{ {{T_R} \ge {t_a}} \right\} \sum\limits_{t = {t_a}}^\infty  {{\bbE_{{t_a}}}\left\{ {\left. {{\bbE_{{t_a}}}\left\{ {\left. {\sum\limits_{m = 1}^M {\max \left\{ {\zeta _m^{(t)},0} \right\}} \bone\left\{ {{T_R}\ \ge t} \right\}} \right|{{\cal F}_{t - 1}}} \right\}} \right|{{\cal F}_{{t_a-1}}}} \right\}} \\ \label{explain_6_1}
& = \bone\left\{ {{T_R} \ge {t_a}} \right\} \sum\limits_{t = {t_a}}^\infty  {{\bbE_{{t_a}}}\left\{ {\left. {\bone\left\{ {{T_R} \ge t} \right\}\sum\limits_{m = 1}^M {{\bbE_{t_a}}\left\{ {\max \left\{ {\zeta _m^{(t)},0} \right\}} \right\}} } \right|{{\cal F}_{{t_a-1}}}} \right\}},
\end{align}
where (\ref{explain_1_0}) comes from the fact that $\bone\{ {{T_R} \ge {t_a}} \}$ is ${\cal F}_{t_a-1}$-measurable,  (\ref{explain_1_1}) is due to the monotone convergence theorem, and (\ref{explain_5_1}) is obtained by the Tower's property and the fact that ${\cal F}_{t_a-1} \subseteq {\cal F}_{t-1} $ when $t \ge t_a$. (\ref{explain_6_1}) holds due to the fact that $\{\zeta _m^{(t)}\}_t$ is an independent sequence for each $m$ and $\bone\{ {{T_R} \ge t} \}$ is ${\cal F}_{t-1}$-measurable.
%  (\ref{explain_2}) is due to the fact that $x_m^{(t)}$ follows the postattack distribution when $t \ge t_a$, and (\ref{explain_3}) is from (\ref{expected_max_zeta_1_lb}) in Lemma \ref{Lemma_bounds_on_expectations}.

By employing (\ref{expected_max_zeta_1_lb}) in Lemma \ref{Lemma_bounds_on_expectations} and (\ref{explain_6_1}), we can obtain
\begin{align} \notag
& {\bbE_{{t_a}}}\left\{ \bone\left\{ {{T_R} \ge {t_a}} \right\}{\left. {\sum\limits_{t = {t_a}}^{{T_R}} {\sum\limits_{m = 1}^M {\max \left\{ {\zeta _m^{(t)},0} \right\}} } } \right|{{\cal F}_{{t_a-1}}}} \right\} \\ \notag 
& \ge \bone\left\{ {{T_R} \ge {t_a}} \right\} \sum\limits_{t = {t_a}}^\infty  {{\bbE_{{t_a}}}\left\{ {\left. {\bone\left\{ {{T_R} \ge t} \right\}} \right|{{\cal F}_{{t_a-1}}}} \right\}} \\ \notag
& \qquad \times \sum\limits_{m = 1}^M {\frac{{\rho _L^2}}{{2{\sigma ^2}}}\left\{ {{\rm{erf}}\left( {\frac{{2{\rho _U}}}{{\sqrt 2 \sigma {{\left\| {{{\bf{p}}_m}} \right\|}_2}}}} \right) - {\rm{erf}}\left( {\frac{{{\rho _L} + {\rho _U}}}{{\sqrt 2 \sigma {{\left\| {{{\bf{p}}_m}} \right\|}_2}}}} \right)} \right\}} \\ \notag
%& = {\bbE_{{t_a}}}\left\{ \bone\left\{ {{T_R} \ge {t_a}} \right\} {\left. {\sum\limits_{t = {t_a}}^\infty  {\bone\left\{ {{T_R} } \ge t \right\}}  } \right|{{\cal F}_{{t_a-1}}}} \right\} \\  \label{explain_4}
%&  \qquad \times \sum\limits_{m = 1}^M {\frac{{\rho _L^2}}{{2{\sigma ^2}}}\left\{ {{\rm{erf}}\left( {\frac{{2{\rho _U}}}{{\sqrt 2 \sigma {{\left\| {{{\bf{p}}_m}} \right\|}_2}}}} \right) - {\rm{erf}}\left( {\frac{{{\rho _L} + {\rho _U}}}{{\sqrt 2 \sigma {{\left\| {{{\bf{p}}_m}} \right\|}_2}}}} \right)} \right\}} \\ \notag
& = {\bbE_{{t_a}}}\left\{ \bone\left\{ {{T_R} \ge {t_a}} \right\} {\left. {\sum\limits_{t = 1}^\infty  {\bone\left\{ {\left( {{T_R} - {t_a} + 1} \right) } \ge t \right\}}  } \right|{{\cal F}_{{t_a-1}}}} \right\} \\ \label{explain_3_1}
& \qquad \times \sum\limits_{m = 1}^M {\frac{{\rho _L^2}}{{2{\sigma ^2}}}\left\{ {{\rm{erf}}\left( {\frac{{2{\rho _U}}}{{\sqrt 2 \sigma {{\left\| {{{\bf{p}}_m}} \right\|}_2}}}} \right) - {\rm{erf}}\left( {\frac{{{\rho _L} + {\rho _U}}}{{\sqrt 2 \sigma {{\left\| {{{\bf{p}}_m}} \right\|}_2}}}} \right)} \right\}} \\  \notag
& = {\bbE_{{t_a}}}\left\{ {\left. {\left( {{T_R} - {t_a} + 1} \right)\bone\left\{ {{T_R} \ge {t_a}} \right\}} \right|{{\cal F}_{{t_a-1}}}} \right\}\sum\limits_{m = 1}^M {\frac{{\rho _L^2}}{{2{\sigma ^2}}}\left\{ {{\rm{erf}}\left( {\frac{{2{\rho _U}}}{{\sqrt 2 \sigma {{\left\| {{{\bf{p}}_m}} \right\|}_2}}}} \right) - {\rm{erf}}\left( {\frac{{{\rho _L} + {\rho _U}}}{{\sqrt 2 \sigma {{\left\| {{{\bf{p}}_m}} \right\|}_2}}}} \right)} \right\}} \\ \label{E_t_a_lb_1}
& = {\bbE_{{t_a}}}\left\{ {\left. {{{\left( {{T_R} - {t_a} + 1} \right)}^ + }} \right|{{\cal F}_{{t_a-1}}}} \right\}\sum\limits_{m = 1}^M {\frac{{\rho _L^2}}{{2{\sigma ^2}}}\left\{ {{\rm{erf}}\left( {\frac{{2{\rho _U}}}{{\sqrt 2 \sigma {{\left\| {{{\bf{p}}_m}} \right\|}_2}}}} \right) - {\rm{erf}}\left( {\frac{{{\rho _L} + {\rho _U}}}{{\sqrt 2 \sigma {{\left\| {{{\bf{p}}_m}} \right\|}_2}}}} \right)} \right\}},
\end{align}
where (\ref{explain_3_1}) is due to the monotone convergence theorem and the fact that $\bone\{ {{T_R} \ge {t_a}} \}$ is ${\cal F}_{t_a-1}$-measurable, which completes the proof for the claim in (\ref{claim_1}).

%\begin{equation}
%{\omega ^{({T_C})}} = h + \Delta 
%\end{equation}
%implies
%\begin{align} \notag
%{\omega ^{({T_C})}}\bone\left\{ {{T_C} \ge {t_a}} \right\} & = h\bone\left\{ {{T_C} \ge {t_a}} \right\} + \Delta \bone\left\{ {{T_C} \ge {t_a}} \right\} \\
%& \le h + \bone\left\{ {{T_C} \ge {t_a}} \right\} \sum\limits_{m = 1}^M {\max \left\{ {\zeta _m^{(T_C)},0} \right\}},
%\end{align}
%by employing (\ref{ub_Delta_1}).

Since ${\max \{ {\zeta _m^{(t)},0} \}} \ge 0$,  by employing (\ref{relaxed_GCUSUM}), we know that ${\omega ^{({T_R})}}\bone\{ {{T_R} \ge {t_a}} \}$ can be bounded from below as per
\begin{align} \notag
{\omega ^{({T_R})}}\bone\left\{ {{T_R} \ge {t_a}} \right\} & = \bone\left\{ {{T_R} \ge {t_a}} \right\}\sum\limits_{t = 1}^{{T_R}} {\sum\limits_{m = 1}^M {\max \left\{ {\zeta _m^{(t)},0} \right\}} } \\ \notag
& = \bone\left\{ {{T_R} \ge {t_a}} \right\}\sum\limits_{t = 1}^{\left( {{t_a} - 1} \right)} {\sum\limits_{m = 1}^M {\max \left\{ {\zeta _m^{(t)},0} \right\}} }  + \bone\left\{ {{T_R} \ge {t_a}} \right\}\sum\limits_{t = {t_a}}^{{T_R}} {\sum\limits_{m = 1}^M {\max \left\{ {\zeta _m^{(t)},0} \right\}} } \\ \label{lb_on_omega}
& \ge \bone\left\{ {{T_R} \ge {t_a}} \right\}\sum\limits_{t = {t_a}}^{{T_R}} {\sum\limits_{m = 1}^M {\max \left\{ {\zeta _m^{(t)},0} \right\}} }, 
\end{align}
where we define $\sum_{t = {t_1}}^{{t_2}}  = 0$ if $t_2 < t_1$. Therefore, by employing the claim in (\ref{claim_1}) and (\ref{lb_on_omega}), we have
\begin{equation} 
{\bbE_{{t_a}}}\left\{ {\left. {{{\left( {{T_R} - {t_a} + 1} \right)}^ + }} \right|{{\cal F}_{{t_a} - 1}}} \right\} \le \frac{{{\bbE_{{t_a}}}\left\{ {\left. {{\omega ^{({T_R})}}\bone\left\{ {{T_R} \ge {t_a}} \right\}} \right|{{\cal F}_{{t_a} - 1}}} \right\}}}{{\sum\limits_{m = 1}^M {\frac{{\rho _L^2}}{{2{\sigma ^2}}}\left\{ {{\rm{erf}}\left( {\frac{{2{\rho _U}}}{{\sqrt 2 \sigma {{\left\| {{{\bf{p}}_m}} \right\|}_2}}}} \right) - {\rm{erf}}\left( {\frac{{{\rho _L} + {\rho _U}}}{{\sqrt 2 \sigma {{\left\| {{{\bf{p}}_m}} \right\|}_2}}}} \right)} \right\}} }}
\end{equation}
which completes the proof for Lemma \ref{Lemma_Conditional_Expectation_ub}.	
\end{IEEEproof}
	
It is seen from (\ref{lorden_formulation_TR}) that the computation of the worst-case expected detection delay of the proposed RGCUSUM requires taking the supremum of $J_{t_a}\left( T_R \right)$  over $t_a$. If the stopping time $T_R$ in (\ref{GCUSUM_simplified}) is an equalizer rule, that is, $J_{t_a}\left( T_R \right)$ is  constant over $t_a$, then the computation of the  worst-case expected detection delay can be greatly simplified, since we can get rid of the supremum in (\ref{lorden_formulation_TR}) when we compute the worst-case expected detection delay. Similar to the conclusion that the stopping time employed in the CUSUM test is an equalizer rule for the classic Lorden's problem \cite{poor2009quickest}, we have the following lemma.
\begin{lemma} \label{Lemma_equalizer_rule}
	The stopping time $T_R$ employed in the RGCUSUM in (\ref{GCUSUM_simplified}) achieves an equalizer rule, i.e.,  
\begin{equation} \label{Lemma_equalizer_rule_J_equal_J_1}
J\left( T_R \right) = \mathop {\sup }\limits_{t_a}  {J_{{t_a}}}\left( {{T_R}} \right) = {J_1}\left( {{T_R}} \right).
\end{equation}
\end{lemma}
\begin{IEEEproof}
It is seen from (\ref{relaxed_GCUSUM}) that for any given $K_1 \le K$, we have
\begin{align} \notag
{\omega ^{(K)}}
%& = \sum\limits_{t = 1}^K {\sum\limits_{m = 1}^M {\max \left\{ {\zeta _m^{(t)},0} \right\}} } \\  \notag
&= \sum\limits_{t = 1}^{{K_1}} {\sum\limits_{m = 1}^M {\max \left\{ {\zeta _m^{(t)},0} \right\}} }  + \sum\limits_{t = {K_1} + 1}^K {\sum\limits_{m = 1}^M {\max \left\{ {\zeta _m^{(t)},0} \right\}} } \\
& = {\omega ^{({K_1})}} + \sum\limits_{t = {K_1} + 1}^K {\sum\limits_{m = 1}^M {\max \left\{ {\zeta _m^{(t)},0} \right\}} },
\end{align}
which implies that ${\omega ^{(K)}}$ is an increasing function of ${\omega ^{(K_1)}}$ for fixed $\{ {\bf x}^{(K_1+1)}, {\bf x}^{(K_1+2)}, ..., {\bf x}^{(K)} \}$.
%$\{ \{ x_{m}^{(K_1+1)}\}_{m=1}^M, \{ x_{m}^{(K_1+2)}\}_{m=1}^M,...,\{ x_{m}^{(K)}\}_{m=1}^M \}$. 
Thus, on the event $\{T_R \ge t_a\}$, it is seen from (\ref{stopping_time_simplified}) that $T_R$ is a nonincreasing function of ${\omega ^{(t_a-1)}}$. In light of this, the supremum of  ${\bbE_{{t_a}}}\{ { {{{( {{T_R} - {t_a} + 1} )}^ + }} |{{\cal F}_{{t_a} - 1}}} \}$ over ${{\cal F}_{{t_a} - 1}}$ is achieved at ${\omega ^{(t_a-1)}}=0$, since ${\omega ^{(t_a-1)}} \ge 0$ and the event $\{{\omega ^{(t_a-1)}}=0\} \in {\cal F}_{t_a-1}$. As a result,
\begin{align} \notag
{J_{{t_a}}}\left( {{T_R}} \right) & = {\esssup _{{{\cal F}_{{t_a} - 1}}}}{\bbE_{{t_a}}}\left\{ {\left. {{{\left( {{T_R} - {t_a} + 1} \right)}^ + }} \right|{{\cal F}_{{t_a} - 1}}} \right\}\\  \notag
& = {\bbE_{{t_a}}}\left\{ { {{{ {{T_R} - {t_a} + 1} } }} \big|{\omega ^{({t_a} - 1)}} = 0} \right\}\\ \label{J_t_a_simplified}
& ={\bbE_{{t_a}}}\left\{  {{T_R} - {t_a} + 1} \Big| {\max \left\{ {\zeta _m^{(t)},0} \right\} = 0}, \;\forall m \text{ and }  \forall t = 1,2,...{t_a} - 1 \right\},
\end{align}
where (\ref{J_t_a_simplified}) is due to the fact that ${\omega ^{({t_a} - 1)}} = \sum_{t = 1}^{{t_a} - 1} {\sum_{m = 1}^M {\max \{ {\zeta _m^{(t)},0} \}} } $ and $\max \{ {\zeta _m^{(t)},0} \} \ge 0$. Therefore, the event $\{{\omega ^{({t_a} - 1)}} = 0\}$ is equivalent to the event $\{{\max \{ {\zeta _m^{(t)},0} \} = 0}, \;\forall m \text{ and }  \forall t = 1,2,...{t_a} - 1\}$.

As discussed before, for two distinct attack time instants $t_a$ and $t_a'$, the corresponding ${\tilde x_m^{(t_a +t)}}$ and ${\tilde x_m^{(t_a' +t)}}$ are identically distributed for any $t \ge 0$, and hence, ${\zeta_m^{(t_a +t)}}$ and ${\zeta_m^{(t_a' +t)}}$ are identically distributed for any $t \ge 0$ according to the definition of ${\zeta_m^{(t)}}$ in (\ref{tilde_zeta_m_t_define}). As a result, it follows from (\ref{stopping_time_simplified}) and (\ref{J_t_a_simplified}) in turn that
\begin{equation} 
{J_{{t_a}}}\left( {{T_R}} \right) = {J_{{1}}}\left( {{T_R}} \right), \; \forall \; t_a \ge 1,
\end{equation}
which implies
\begin{equation} 
J\left( T_R \right) = \mathop {\sup }\limits_{t_a}  {J_{{t_a}}}\left( {{T_R}} \right) = {J_1}\left( {{T_R}} \right),
\end{equation}
which completes the proof.
\end{IEEEproof}

It is worth mentioning that the equalizer rule property of the CUSUM test for the classic Lorden's problem \cite{poor2009quickest} relies on the fact that the pre- and  post-change observation sequences are both i.i.d. with perfectly known distributions. However, in the problem considered in this paper, the pre- and post-attack observation sequences are not i.i.d., and moreover, there are unknown parameters in their distributions. Lemma \ref{Lemma_equalizer_rule} demonstrates that the proposed RGCUSUM test still has the equalizer rule property for the problem considered in this paper, and therefore, we can just compute  ${J_1}( {{T_R}} )$ instead of $J( T_R )$ to scrutinize the worst-case expected detection delay of the proposed RGCUSUM. We have the following theorem regarding the worst-case expected detection delay of the proposed RGCUSUM.
%The equalizer rule property of the proposed RGCUSUM test for the considered problem still holds, even though the pre-attack and the post-attack observation sequences are not i.i.d. in the considered problem, and there are unknown parameters in their distribution functions. 
%As demonstrated by Lemma \ref{Lemma_equalizer_rule}, we can just compute  ${J_1}( {{T_R}} )$ instead of $J( T_R )$ to scrutinize the worst-case expected detection delay of the proposed RGCUSUM. We have the following theorem regarding the worst-case expected detection delay of the proposed RGCUSUM.

\begin{theorem} \label{Theorem_Detection_Delay}
	By employing Wald's approximations \cite{tartakovsky2014sequential}, i.e., ignoring the expectation of the overshoots in the presence of attacks, for any given $h$, the worst-case expected detection delay of the RGCUSUM can be bounded from above as per
	\begin{equation}
%	J\left( {{T_R}} \right) \le \frac{{h + \sum\limits_{m = 1}^M {\left\{ {\frac{{{\rho _L} + {\rho _U}}}{\sigma }\sqrt {\frac{2}{\pi }} {{\left\| {{{\bf{p}}_m}} \right\|}_2}{e^{ - \frac{{\rho _L^2}}{{2{\sigma ^2}\left\| {{{\bf{p}}_m}} \right\|_2^2}}}} + \frac{{\left( {{\rho _L} + {\rho _U}} \right){\rho _U}}}{{{\sigma ^2}}}\left[ {1 - 2\Phi \left( { - \frac{{{\rho _U}}}{{\sigma {{\left\| {{{\bf{p}}_m}} \right\|}_2}}}} \right)} \right] + \frac{{\rho _U^2}}{{2{\sigma ^2}}} + \frac{1}{2}\left\| {{{\bf{p}}_m}} \right\|_2^2} \right\}} }}{{\sum\limits_{m = 1}^M {\frac{{\rho _L^2}}{{2{\sigma ^2}}}\left\{ {{\rm{erf}}\left( {\frac{{2{\rho _U}}}{{\sqrt 2 \sigma {{\left\| {{{\bf{p}}_m}} \right\|}_2}}}} \right) - {\rm{erf}}\left( {\frac{{{\rho _L} + {\rho _U}}}{{\sqrt 2 \sigma {{\left\| {{{\bf{p}}_m}} \right\|}_2}}}} \right)} \right\}} }}.
J\left( {{T_R}} \right) \le h{\left\{ {\sum\limits_{m = 1}^M {\frac{{\rho _L^2}}{{2{\sigma ^2}}}\left[ {{\rm{erf}}\left( {\frac{{2{\rho _U}}}{{\sqrt 2 \sigma {{\left\| {{{\bf{p}}_m}} \right\|}_2}}}} \right) - {\rm{erf}}\left( {\frac{{{\rho _L} + {\rho _U}}}{{\sqrt 2 \sigma {{\left\| {{{\bf{p}}_m}} \right\|}_2}}}} \right)} \right]} } \right\}^{ - 1}}.
	\end{equation}
\end{theorem}
\begin{IEEEproof}
At time instant $T_R$, we have
\begin{equation}
{\omega ^{({T_R})}} =  h  + \Delta,
\end{equation}
where $\Delta$ is the overshoot. This implies that for any given $t_a$,
%which implies
\begin{equation} \label{proof_theorem_temp1}
%{\omega ^{({T_R})}}\bone\left\{ {{T_R} \ge {t_a}} \right\} \approx h\bone\left\{ {{T_R} \ge {t_a}} \right\}.
{\omega ^{({T_R})}}\bone\left\{ {{T_R} \ge {t_a}} \right\} =  \left(h  + \Delta \right)\bone\left\{ {{T_R} \ge {t_a}} \right\} \le h  + \Delta.
\end{equation}
It is seen from Lemma \ref{Lemma_equalizer_rule} that when we calculate $J(T_R)$, we can just set $t_a = 1$ without loss of generality, and therefore,
\begin{align} \notag
{\bbE_{{t_a}}}\left\{ {\left. {{\omega ^{({T_R})}}\bone\left\{ {{T_R} \ge {t_a}} \right\}} \right|{{\cal F}_{{t_a} - 1}}} \right\} & = {\bbE}_1\left\{ {{\omega ^{({T_R})}}\bone\left\{ {{T_R} \ge {1}} \right\}} \right\}\\ \notag
	& \le h + {\bbE}_1\left\{ \Delta \right\} \\ \label{E_omega_ub_1}
%	& \approx {\bbE}_1\left\{ {h\bone\left\{ {{T_R} \ge {t_a}} \right\}  } \right\}\\ \label{E_omega_ub_1}
	& \approx h 
\end{align}
by employing (\ref{proof_theorem_temp1}) and Wald's approximations \cite{tartakovsky2014sequential}, i.e., ignoring the expectation of the overshoots in the presence of attacks.
%where (\ref{E_omega_ub_1}) holds due to the fact that $\Delta \ge 0$. By employing (\ref{Upper_Bound_on_Overshoot}), (\ref{E_omega_ub_1}) and Lemma \ref{Lemma_Conditional_Expectation_ub}, we can obtain
Furthermore, by employing (\ref{Define_J_t_a_TR}), Lemma \ref{Lemma_Conditional_Expectation_ub} and Lemma \ref{Lemma_equalizer_rule}, we can obtain
	\begin{equation}
	J\left( {{T_R}} \right) =J_1\left( {{T_R}} \right)  \le h{\left\{ {\sum\limits_{m = 1}^M {\frac{{\rho _L^2}}{{2{\sigma ^2}}}\left[ {{\rm{erf}}\left( {\frac{{2{\rho _U}}}{{\sqrt 2 \sigma {{\left\| {{{\bf{p}}_m}} \right\|}_2}}}} \right) - {\rm{erf}}\left( {\frac{{{\rho _L} + {\rho _U}}}{{\sqrt 2 \sigma {{\left\| {{{\bf{p}}_m}} \right\|}_2}}}} \right)} \right]} } \right\}^{ - 1}},
%	\frac{{h + \sum\limits_{m = 1}^M {\left\{ {\frac{{{\rho _L} + {\rho _U}}}{\sigma }\sqrt {\frac{2}{\pi }} {{\left\| {{{\bf{p}}_m}} \right\|}_2}{e^{ - \frac{{\rho _L^2}}{{2{\sigma ^2}\left\| {{{\bf{p}}_m}} \right\|_2^2}}}} + \frac{{\left( {{\rho _L} + {\rho _U}} \right){\rho _U}}}{{{\sigma ^2}}}\left[ {1 - 2\Phi \left( { - \frac{{{\rho _U}}}{{\sigma {{\left\| {{{\bf{p}}_m}} \right\|}_2}}}} \right)} \right] + \frac{{\rho _U^2}}{{2{\sigma ^2}}} + \frac{1}{2}\left\| {{{\bf{p}}_m}} \right\|_2^2} \right\}} }}{{\sum\limits_{m = 1}^M {\frac{{\rho _L^2}}{{2{\sigma ^2}}}\left\{ {{\rm{erf}}\left( {\frac{{2{\rho _U}}}{{\sqrt 2 \sigma {{\left\| {{{\bf{p}}_m}} \right\|}_2}}}} \right) - {\rm{erf}}\left( {\frac{{{\rho _L} + {\rho _U}}}{{\sqrt 2 \sigma {{\left\| {{{\bf{p}}_m}} \right\|}_2}}}} \right)} \right\}} }},
	\end{equation}
	which completes the proof.
\end{IEEEproof}

As illustrated in Theorem \ref{Theorem_Detection_Delay}, the worst-case expected detection delay of the RGCUSUM can be bounded by a term which is proportional to the threshold $h$.
It is worth mentioning that Wald's approximations are employed in Theorem \ref{Theorem_Detection_Delay}, which implicitly assumes that the expectation of the overshoot should be negligibly small when compared to the threshold $h$. In Section \ref{Section_Application_Smart_Grid}, we will numerically examine the validity of the Wald's approximations in some practical cases, and the results show that the Wald's approximations are valid when $h$ is large.

{\it Remark:} It is worth pointing out that it seems impossible to conduct a similar performance analysis for the GCUSUM defined in (\ref{GCUSUM_test}). The reasons are mainly twofold. On one hand, unlike $\zeta_m^{(t)}$ defined in (\ref{tilde_zeta_m_t_define}), there is no closed-from expression for $v_t$ in (\ref{Define_w_t}), which hinders us from deriving the statistical characterization of $v_t$. On the other hand, as demonstrated in (\ref{RGCUSUM_recursive}), the test statistic $\omega^{(K)}$ of the RGCUSUM can be written as a linear function of $\omega^{(K-1)}$ in its recursive expression, which facilitates the performance analysis of the RGCUSUM. However, it is seen from (\ref{recursive_GCUSUM}) that $V_K$ is a nonlinear function of $V_{K-1}$, and hence, the performance analysis procedure for the RGCUSUM cannot be carried out for the GCUSUM.

\section{Simulation Results}
\label{Section_Application_Smart_Grid}

\begin{figure}[htb]
	%\vspace{-0.10in}
	\vspace{0.028in}
	\centerline{
		\includegraphics[width=0.56\textwidth]{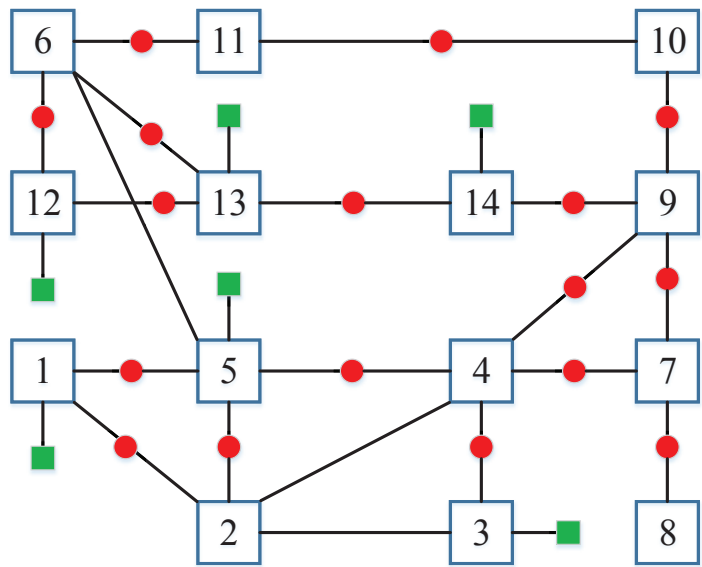}
	}
	\vspace{0.15in}
	\caption{IEEE 14-bus power system. The red circles on the branches denote power-flow measurements, and the green squares represent the power injection measurements.}
	\label{Fig_IEEE_14_bus}
\end{figure}

%\begin{figure}[htb]
%	%\vspace{-0.10in}
%	\vspace{0.028in}
%	\centerline{
%		%\includegraphics[width=0.46\textwidth]{20151019_surf_everypoint_nonidentifiable_Xalpha_beta10_cut.pdf}
%		\includegraphics[width=0.46\textwidth]{Overshoot_H0_cut.eps}
%	}
%	\vspace{0.15in}
%	\caption{Average ratio of overshoot to threshold in the absence of attacks.}
%	\label{Fig_overshoot_H0}
%\end{figure}

In this section, we consider applying the proposed RGCUSUM algorithm to a dynamic smart grid system, and we numerically evaluate its performance.

Simulations are performed on an IEEE-14 bus power system which is illustrated in Fig. \ref{Fig_IEEE_14_bus}, and the measurement matrix ${\bf{H}}$ in (\ref{regression_model}) is determined accordingly for the dynamic DC model of the power system in Fig. \ref{Fig_IEEE_14_bus}. The initial state of the power system is defined in the MATPOWER ``case14'' \cite{zimmerman2010matpower}. MATPOWER is a MATLAB simulation toolbox for solving optimal power flow problems, which can provide realistic power flow data for different test systems that are widely considered in research-oriented studies as well as in practice.
We assume that the resistive load at bus 3 decreases by 100 watts per time instant, while the resistive loads at bus 5 and bus 11 both increase by 100 watts per time instant. As such, the state variables of the smart grid change accordingly over time.  It is worth mentioning that due to the nonlinear relationship between the state parameter vector $\btheta^{(t)}$ and the resistive loads at the buses, $\btheta^{(t)}$ cannot be explicitly expressed as a function of the resistive loads at the buses.  To this end,  for each $t$, we resort to some numerical method to obtain the time-varying state parameter vector $\btheta^{(t)}$  for the simulations. To be specific, the MATPOWER ``case14'' DC power-flow algorithm is employed to generate $\btheta^{(t)}$ for each $t$ by taking the dynamic resistive loads  into account.

\subsection{Performance Evaluation}
\begin{figure}[htb]
	%\vspace{-0.10in}
	\vspace{0.028in}
	\centerline{
		\includegraphics[width=0.66\textwidth]{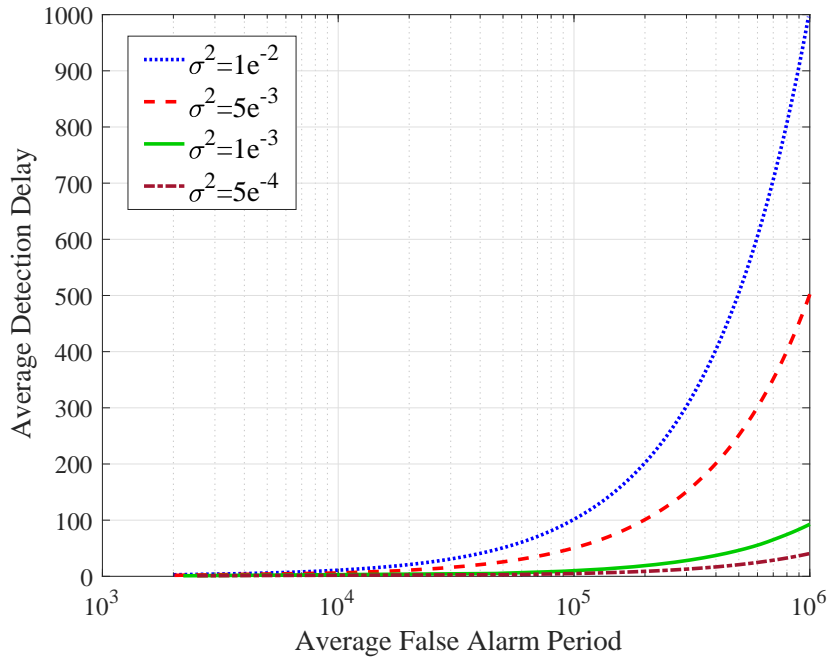}
	}
	\vspace{0.15in}
	\caption{Performance of the proposed RGCUSUM algorithm.}
	\label{Fig_Performance_RGCUSUM}
\end{figure}

The performance of the proposed RGCUSUM algorithm is illustrated in  Fig. \ref{Fig_Performance_RGCUSUM} where the variance of noise $\sigma^2=0.01,0.005,0.001$, and $0.0005$, respectively. In the simulation, $\rho_L$ and $\rho_U$ are set to be $0.025$ and $100$, respectively. The number of Monte Carlo runs is $300$.  Although the proposed RGCUSUM can be applied to the cases where the FDIA ${\bf a}^{(t)}$ is time-varying, for simplicity, ${\bf a}^{(t)}$ is set to be a constant vector in the simulations, which is 
\begin{align} \notag
{\bf a}^{(t)}& =[  -2.629, -2.704, 2.781, 2.923, 0.516, -0.936, 1.969, -3.938, -0.033, 0, -0.483, -0.033, \\  \label{constant_attack}
& \; -1.934, 1.934, -1.934, 4.259, 2.842, 0.110, 1.314, -0.520, 2.195, -0.046, 1.778]^T, \; \forall  t \ge 1.
\end{align}
As shown in Fig. \ref{Fig_Performance_RGCUSUM},  for a given average detection delay, the average false alarm period of the RGCUSUM increases as $\sigma^2$ decreases. These numerical results agree with
our theory presented in Section \ref{Section_Performance_Analysis}. To be specific, it is seen from (\ref{tilde_zeta_m_t_define}) and (\ref{tilde_x_distribution_1}) that in the region where $\sigma^2$ is small, the expectation of $\zeta_m^{(t)}$ is mainly determined by the injected data ${\bf a}^{(t)}$ in the presence of attacks, especially when ${\bf p}_m^T{\bf a}^{(t)}$ is large. In light of this, we know from (\ref{stopping_time_simplified}) that for a given threshold $h$, the change of $\sigma^2$ cannot significantly affect the average detection delay of the RGCUSUM in the presence of attacks in the region where $\sigma^2$ is small. On the other hand, as demonstrated by (\ref{tilde_zeta_m_t_define}) and (\ref{tilde_x_distribution_0}), the decrease of $\sigma^2$ in the absence of attacks brings about a relatively large decrease in the expectation of $\zeta_m^{(t)}$, and therefore, it is seen from (\ref{stopping_time_simplified}) that in the absence of attacks, the average false alarm period of the RGCUSUM   increases as $\sigma^2$ decreases. To this end, it is not surprising that for a given average detection delay, the average false alarm period of the RGCUSUM algorithm increases as the variance of noise $\sigma^2$ decreases in  Fig. \ref{Fig_Performance_RGCUSUM}.

\begin{figure}[htb]
	%\vspace{-0.10in}
	\vspace{0.028in}
	\centerline{
		\includegraphics[width=0.66\textwidth]{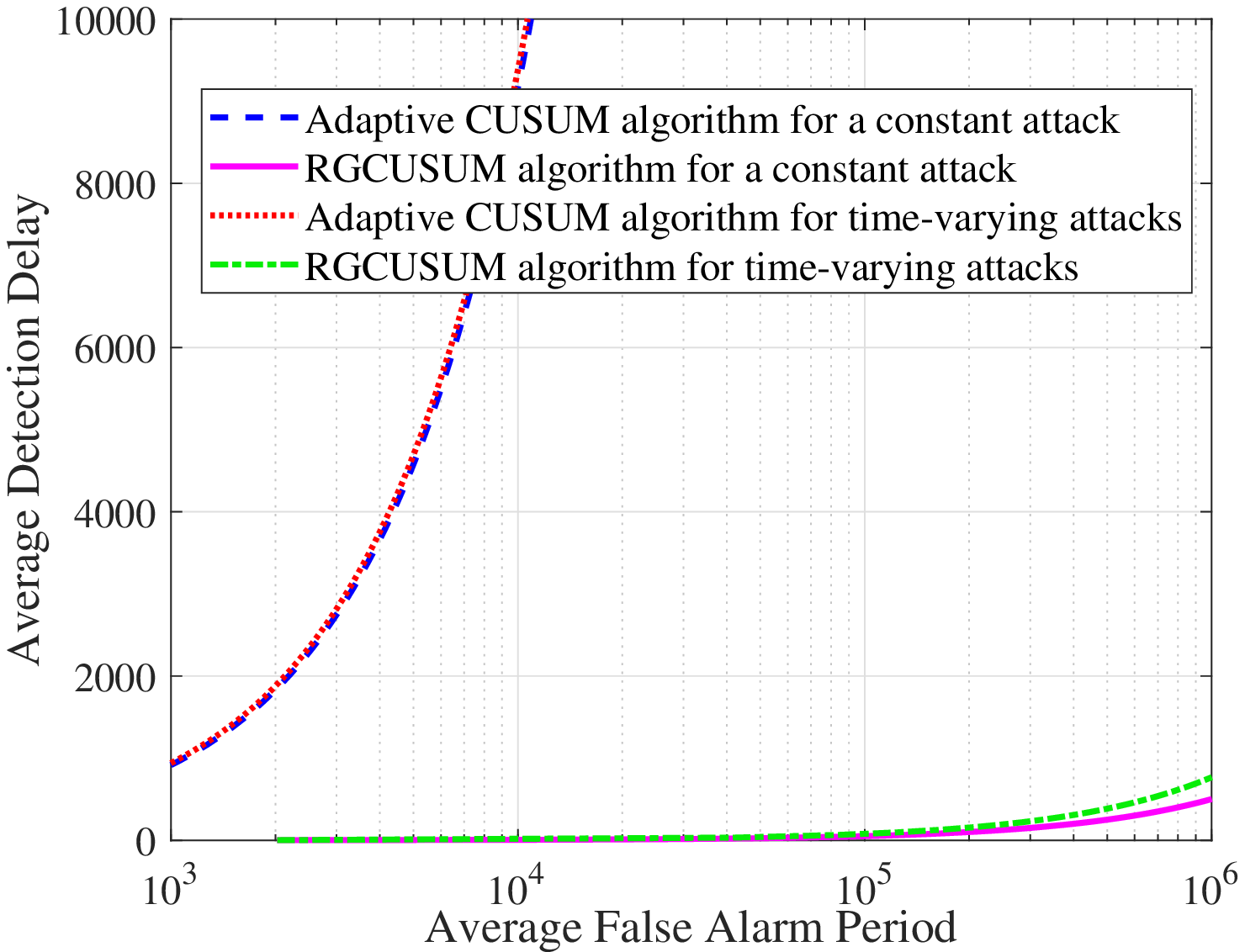}
	}
	\vspace{0.15in}
	\caption{Performance comparison between the proposed RGCUSUM algorithm and the adaptive CUSUM algorithm.}
	\label{Fig_Performance_Comparison}
\end{figure}

\subsection{Performance Comparison}

In this subsection, we compare the performance of the proposed RGCUSUM to that of  existing approaches for the same simulation setup described at the beginning of this section. As explained in Section \ref{Subsection_Related_Work}, most of existing approaches cannot be applied to the simulation setup considered in this paper. For example, the Kalman filter based algorithm proposed in \cite{kurt2018distributed} requires that the evolution of the state parameter vector $\btheta^{(t)}$ with $t$ follows a fixed linear model, and the linear matrix must be known to the control center. However,  for the simulation setup considered in this section, the evolution of the state parameter vector $\btheta^{(t)}$ with $t$ does not follow a linear model, and moreover, it seems impossible to explicitly characterize this evolution. Thus, the decision statistic proposed in \cite{kurt2018distributed} cannot be evaluated here. Likewise, the decision statistic proposed in \cite{huang2016real} also cannot be evaluated for the simulation setup considered in this section, since the covariance matrix of the residual is singular.

In the following, we compare the performance of the RGCUSUM with that of a representative approach, called adaptive CUSUM algorithm proposed in \cite{huang2011defending}. The numerical comparison is illustrated in Fig. \ref{Fig_Performance_Comparison}. In the simulation, the variance of noise $\sigma^2=0.005$, and $\rho_L$ and $\rho_U$ are set to be $0.025$ and $100$, respectively. The number of Monte Carlo runs is $300$. In the comparison, we consider two scenarios. In one scenario,  ${\bf a}^{(t)}$ is set to be a constant vector as defined in (\ref{constant_attack}). In the other scenario, ${\bf a}^{(t)}$ is time-varying, which are defined as, $\forall  r=0,1,2,...$,
\begin{align} \notag
{{\mathbf{a}}^{(3r+1)}}  = &   [ 0, 6.881,-1.776,-3.067,0.747,0.949,0.545,-1.090,-0.249,0,-0.498,-0.249,  -0.351, \\ \label{time_varying_attack_1}
& \; 0.351,-0.351,-0.395,0,0.395,0,0.264,0.132,0.132,0]^T  \times [1 + (3r+1) \times {10^{ - 6}}],
\end{align}
\begin{align} \notag
{{\mathbf{a}}^{(3r+2)}}  = [& -3.528,-0.375,-0.246,-0.504,1.008,0.125,0,0.250,0.125,0.176,-0.176,0.176, \\ \label{time_varying_attack_2}
& \; 0.199,0, -0.199,0,-0.132,-0.066,-0.066,0]^T  \times [1 + (3r+2) \times {10^{ - 6}}],
\end{align}
\begin{align} \notag
{{\mathbf{a}}^{(3r+3)}}  = [& 3.983,5.254,0,-4.445,0.346,0.578,0.116,-0.231,-0.116,0,-0.231,-0.116,-0.163, \\ \label{time_varying_attack_3}
& \; 0.163,-0.163, -0.184,0,0.184,0,0.122,0.061,0.061,0]^T \times [1 + (3r+3) \times {10^{ - 6}}].
\end{align}
It is worth mentioning that ${{\mathbf{a}}^{(3r+1)}}$, ${{\mathbf{a}}^{(3r+2)}}$ and ${{\mathbf{a}}^{(3r+3)}}$ all lie in the complementary space ${{\cal R}^ \bot({\bf H}) }$ of the column space of $\bf H$, and hence, we can see from (\ref{time_varying_attack_1}), (\ref{time_varying_attack_2}) and (\ref{time_varying_attack_3}) that for $r=0,1,2,...$, ${[{\mathcal{A}^{(3r + 1)}}]^C} = \{ 1,10,17,19,23\} $, ${[{\mathcal{A}^{(3r + 2)}}]^C} = \{ 2,10,17,19,23\} $ and ${[{\mathcal{A}^{(3r + 3)}}]^C} = \{ 3,10,17,19,23\}$, where $[\cA^{(t)}]^C$ denotes the complement of $\cA^{(t)}$ for any set $\cA^{(t)}$ of nonzero elements of $\bmu^{(t)}$.

It is seen from Fig. \ref{Fig_Performance_Comparison} that for a given average false alarm period, the average detection delay of the proposed RGCUSUM algorithm is shorter than that of the adaptive CUSUM algorithm in both scenarios, which implies that the proposed RGCUSUM algorithm can detect the FDIAs more efficiently than the adaptive CUSUM algorithm. This is expected since the adaptive CUSUM algorithm builds on a different model.  To be specific, the adaptive CUSUM algorithm assumes that $\btheta^{(t)}$ is a Gaussian vector with some known mean and covariance. However, in the simulation,  $\btheta^{(t)}$ is an unknown deterministic vector for each $t$. The efficiency loss of the adaptive CUSUM algorithm may be brought about by the model mismatch. Moreover, the adaptive CUSUM algorithm requires that the FDIAs are positive and small, and hence is prone to efficiency loss for large and negative FDIA, which is the case in the simulation. In addition, as illustrated in Fig. \ref{Fig_Performance_Comparison}, the performance of the RGCUSUM algorithm in the scenario where ${\bf a}^{(t)}$ is constant is better than that in the scenario where ${\bf a}^{(t)}$ is time-varying. One possible reason is that for ${\bf a}^{(t)}$ in (\ref{constant_attack}), the corresponding $[\cA^{(t)}]^C = \{10\}$ for all $t$. In light of this, in the scenario where  ${\bf a}^{(t)}$ is defined in (\ref{constant_attack}), the number of attacked meters is more than that in the scenario where ${\bf a}^{(t)}$ is defined in (\ref{time_varying_attack_1}), (\ref{time_varying_attack_2}) and (\ref{time_varying_attack_3}). Hence, it should be easier for the RGCUSUM algorithm to detect the FDIAs in the case where  ${\bf a}^{(t)}$ is defined in (\ref{constant_attack}).

\begin{figure}[htb]
	%\vspace{-0.10in}
	\vspace{0.028in}
	\centerline{
		\includegraphics[width=0.66\textwidth]{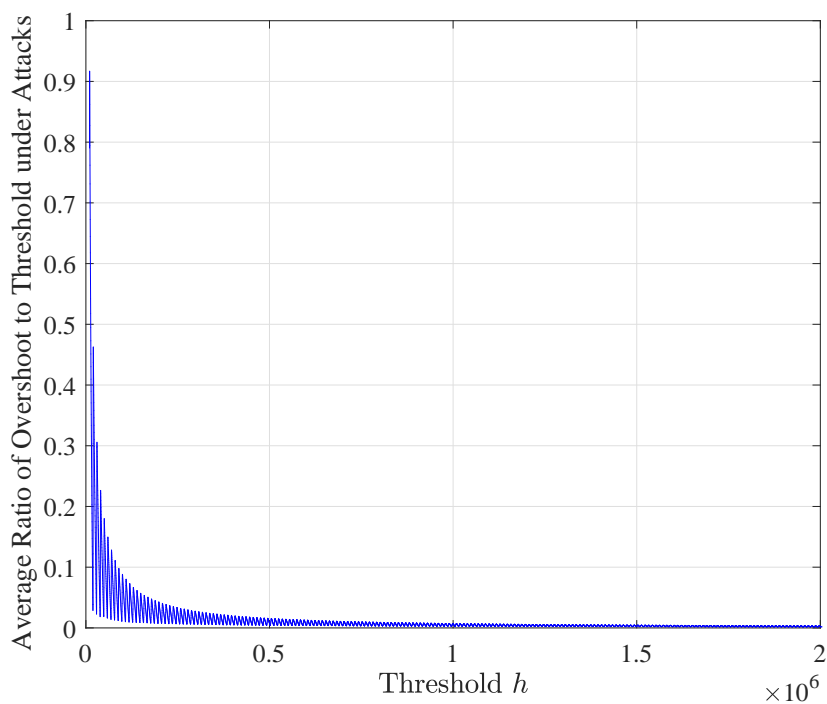}
	}
	\vspace{0.15in}
	\caption{Average ratio of overshoot to threshold in the presence of attacks.}
	\label{Fig_overshoot_H1}
\end{figure}

\subsection{Numerical Verification of Negligible Expected Overshoot}

In the proof of  Theorem \ref{Theorem_Detection_Delay}, the expectation of the overshoot is ignored as shown in (\ref{E_omega_ub_1}), which essentially requires that in the presence of attacks,  the expectation of the overshoot is negligible when compared to the threshold $h$. Here, we numerically scrutinize the average overshoot as a percentage of the threshold $h$ in the presence of attacks. In the simulation as shown in Fig. \ref{Fig_overshoot_H1}, we consider the same case as that considered in Fig. \ref{Fig_Performance_RGCUSUM} where $\sigma^2=0.005$ and the number of Monte Carlo runs is $600$. Fig. \ref{Fig_overshoot_H1} depicts that as the threshold $h$ increases, the average ratio of the overshoot to $h$ decreases to $0$ in the presence of attacks. In addition, for the other cases considered in Fig. \ref{Fig_Performance_RGCUSUM}, we observe similar results that the average ratio of the overshoot to $h$ decreases to $0$ as $h$ grows. Therefore, Wald's approximations employed in Theorem \ref{Theorem_Detection_Delay} are valid when the threshold is sufficiently large.

\section{Conclusions}
\label{Section_Conclusions}

In this paper, we have considered a general problem of sequentially detecting time-varying FDIAs in dynamic linear regression models, which subsumes the problem of sequentially detecting the FDIAs in dynamic smart grid systems when the DC power flow model is adopted. The GCUSUM algorithm and the RGCUSUM have been proposed to solve this problem, and we have shown that the computational complexity of the proposed RGCUSUM algorithm scales linearly with the number of observations. Thus, the proposed RGCUSUM algorithm is amenable to implementation in practice. Furthermore, we also have provided performance analysis of the proposed RGCUSUM algorithm. To be specific, considering Lordon's setup, for any given constraint on the expected false alarm period, a lower bound on the threshold employed in the proposed RGCUSUM algorithm has been derived, which provides a guideline for the design of the proposed RGCUSUM algorithm to achieve the prescribed performance requirement. In addition, for any given threshold employed in the proposed  RGCUSUM algorithm, we also have provided an upper bound on the worst-case expected detection delay. Finally, the performance of the proposed RGCUSUM algorithm has been numerically studied based on an IEEE standard power system under FDIAs.

\bibliographystyle{IEEEtran}
\bibliography{StochasticEncryption}

\end{document}